\documentclass[twocolumn,aps,nofootinbib,floatfix]{revtex4}
\usepackage{graphicx,subfigure}
\usepackage{multirow}
\usepackage[dvips]{epsfig}
\usepackage{amsmath,amssymb,mathrsfs,wasysym,latexsym}
\usepackage{cases,color}
\newcommand{\dLips}{{\rm d}\mathscr{L}\!\textsl{ips}}
\renewcommand{\Im}{\mathfrak{Im}}
\renewcommand{\Re}{\mathfrak{Re}}
\def\lsim{\raise0.3ex\hbox{$\;<$\kern-0.75em\raise-1.1ex\hbox{$\sim\;$}}}
\def\gsim{\raise0.3ex\hbox{$\;>$\kern-0.75em\raise-1.1ex\hbox{$\sim\;$}}}

\begin{document}
\title{Normal tau polarisation as a sensitive probe of CP violation
       in chargino decay}
\author{Herbi~K.~Dreiner$^1$, Olaf~Kittel$^2$, Anja~Marold$^1$}

\affiliation{$^1$ Bethe Centre for Theoretical Physics \& Physikalisches
                  Institut, Universit\"at Bonn, D-53115 Bonn, Germany\\ 
             $^2$ Departamento de F\'isica Te\'orica y del Cosmos and CAFPE,
                  Universidad de Granada, E-18071 Granada, Spain}
\begin{abstract}
CP violation in the spin-spin correlations in chargino production and
subsequent two-body decay into a tau and a tau-sneutrino is studied at
the ILC. From the normal  
polarisation of the tau, an asymmetry is defined to test the
CP-violating phase of the higgsino mass parameter $\mu$. Asymmetries
of more than $\pm70\%$ are obtained, also in scenarios with heavy
first and second generation sfermions.  Bounds on the statistical
significances of the CP asymmetries are estimated. As a result, the
normal tau polarisation in the chargino decay is one of the most
sensitive probes to constrain or measure the phase $\varphi_\mu$ at
the ILC, motivating further detailed experimental studies.
\end{abstract}
\maketitle
\section{Introduction}\label{sec:intro}

Within the framework of the standard electro-weak model (SM), the
complex Cabbibo-Kobayashi-Maskawa (CKM) matrix is the origin for CP
violation~\cite{CKMref,Amsler:2008zzb,Buras:2005xt}. It
explains all the current laboratory data, but it is not sufficient to
generate the matter-antimatter asymmetry of the
universe~\cite{Csikor:1998eu}. Thus further theories have to
be investigated, that offer new sources of CP
violation~\cite{Riotto:1998bt}. The minimal supersymmetric standard
model (MSSM) is a promising extension of the
SM~\cite{Haber:1984rc}. To prevent the supersymmetric partners of the
known particles from appearing below the LEP and Tevatron energy
scale, there has to be supersymmetry (SUSY) break-down, at
least at the electro-weak energy scale~\cite{Claudson:1983et}. Several
of the supersymmmetric parameters can be complex, including the
higgsino mass parameter $\mu$.

\medskip

Concerning low energy observables, the corresponding SUSY CP phases
lead to T-violating electric dipole moments (EDMs), that already would
be far beyond the experimental
bounds~\cite{Amsler:2008zzb,Commins:1994gv,Baker:2006ts,Griffith:2009zz,Falk:1999tm,Semertzidis:2003iq}.
This constitutes the SUSY CP problem: The SUSY phases have to be
considerably suppressed, unless cancellations appear between different
EDM contributions, or the SUSY spectrum is beyond the TeV
scale~\cite{Semertzidis:2004uu,cancellations1,cancellations2,cancellations3,Choi:2004rf}. 
And indeed, attempts to naturally solve the SUSY CP problem, often
still require a certain amount of tuning among the SUSY masses, phases
and parameters. Recently proposed models are \emph{split
SUSY}~\cite{ArkaniHamed:2004yi}, \emph{inverted hierarchy}
models~\cite{Bagger:1999ty}, and also
\emph{focus point} scenarios that attempt to restore naturalness~\cite{Feng:1999zg}.
These models have also been proposed to solve the SUSY flavour problem, to
ensure proton stability, and to fulfil cosmological bounds, like constraints
on  dark matter or primordial nucleosynthesis. It is clear, that  CP-sensitive
observables outside the low energy sector have to be proposed and measured, in
order to tackle the SUSY CP problem, and to reveal the underlying  SUSY
model.

\medskip
 
Concerning future collider experiments at the LHC~\cite{LHC} and
ILC~\cite{ILC}, SUSY phases alter SUSY particle masses, cross
sections, branching ratios~\cite{Bartl:2002uy,Rolbiecki:2009hk}, and
longitudinal polarisations of final state
fermions~\cite{Gajdosik:2004ed}. Although the SUSY phases can change
these CP-even observables by an order of magnitude or more, only
CP-odd observables are a direct evidence of CP
violation~\cite{Han:2009ra}.  CP-odd rate asymmetries of cross
sections, distributions, or partial decay widths~\cite{Choi:1998ub},
however, usually do not exceed $10\%$, as they require the presence of
absorptive phases, unless they are resonantly
enhanced~\cite{Pilaftsis:1997dr,Choi:2004kq,Accomando:2006ga,Nagashima:2009gm}.
At tree level, larger T-odd and CP-odd observables can be defined with
triple or epsilon products of particle momenta and/or
spins~\cite{Valencia:1994zi,CPreview}.

\medskip 

At the LHC, CP-odd triple product asymmetries have been studied in the decays
of third generation
squarks~\cite{Bartl:2004jr,Bartl:2002hi,Deppisch:2009nj,Ellis:2008hq}, and 
three-body decays of neutralinos which originate from
squarks~\cite{Ellis:2008hq,Langacker:2007ur,MoortgatPick:2009jy}.
Since triple products are not boost invariant, compared to epsilon products,
some of these studies have included boost effects at the
LHC~\cite{Deppisch:2009nj,Ellis:2008hq,Langacker:2007ur,MoortgatPick:2009jy}.
For the ILC, triple product asymmetries have been studied in the production and
decay of neutralinos~\cite{Bartl:2003gr,Choi:2003pq,NEUT2,NEUT3,Kittel:2004rp},
and charginos~\cite{Kittel:2004rp,CHAR2,CHAR3}, also using transversely
polarised beams~\cite{Trans}. The result of these studies is, that the largest
asymmetries of the order of $60\%$ can be obtained if final fermion
polarisations are analysed, like the  normal tau polarisation in  neutralino
decay $\tilde\chi_i^0\to \tilde\tau_k \tau$~\cite{Bartl:2003gr,Choi:2003pq}.

\medskip

Although the experimental reconstruction of taus is much more involved
than those for electrons or muons, tau decays in principle allow for a
measurement of their
polarisation~\cite{Berge:2008wi,Dreiner:1992gt}. The fermion
polarisation contains additional and unique information on the SUSY
couplings~\cite{Boos:2003vf,Perelstein:2008zt}, which will be lost if
only electron or muon momenta are considered for measurements.  Taus
might also be the dominant final state leptons. In particular
in inverted hierarchy models, sleptons of the first and second
generations are heavy, such that electron and muon final states will
be rare in SUSY particle decay chains. We are thus motivated to
analyse the potential of CP-odd effects in chargino production
\begin{equation}
  e^+ + e^- \to {\tilde{\chi}}_i^\pm + {\tilde{\chi}}_j^\mp,
                \quad   i,j=1,2,
\label{charprod}
\end{equation}
with longitudinally polarised beams, and the subsequent two-body decay of one
chargino into a polarised tau
\begin{equation}
  \tilde{\chi}_i^\pm \to \tau^\pm + \tilde{\nu}^{(\ast)}_\tau.
\label{chardec}
\end{equation}
For the chargino decay, we define a CP-odd asymmetry from the 
tau polarisation normal to the plane spanned by the $e^-$-beam and the
  tau momentum. It is highly
sensitive to the CP phase $\varphi_\mu$ of the higgsino mass parameter
$\mu$, which enters the chargino sector.  Since $\varphi_\mu$ is also
the most constrained SUSY CP phase from EDM searches, this asymmetry
is particularly important.  Besides a MSSM scenario with light
sleptons, we thus also analyse the CP asymmetries in an inverted
hierarchy scenario, which relaxes the strong EDM constraints on the
SUSY phases.  Scenarios with an inverted hierarchy are attractive
for our process of chargino production and decay, since
not only is the chargino production cross section enhanced due to
missing destructive sneutrino interference, but the rate of chargino
decay into taus is also amplified.

\medskip

In Section~\ref{sec:formalism}, we present our formalism.  We briefly
review chargino mixing in the complex MSSM, and give the relevant
parts of the interaction Lagrangian.  We calculate the $\tau$ spin
density matrix, the normal $\tau$ polarisation, and present the
corresponding analytical formulae in the spin-density matrix
formalism~\cite{Haber:1994pe}. In Section~\ref{sec:results}, we
present our numerical results.  We summarise and conclude in
Section~\ref{sec:summary}.

\section{Formalism}\label{sec:formalism}

\subsection{Chargino mixing and complex couplings}\label{sub:mixing}

\begin{figure}[t!] 
  \includegraphics[width=0.99\columnwidth]{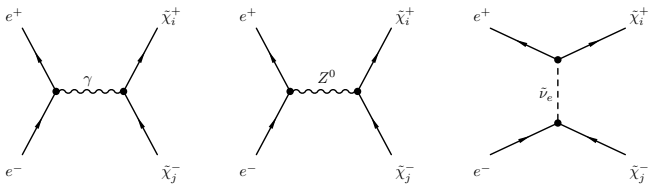} 
  \caption{Feynman diagrams for chargino production.} 
\label{fig:1} 
\end{figure}
In the MSSM, the charged winos $\tilde{W}^\pm$ and higgsinos
$\tilde{H}^\pm$ mix after electro-weak symmetry breaking,
and form the chargino mass eigenstates
$\tilde{\chi}^\pm_{1,2}$. In the $(\tilde W,\tilde H)$ basis, their
mixing is defined by the complex chargino mass
matrix~\cite{Haber:1984rc}
\begin{eqnarray} 
  M_{\tilde\chi}= 
  \left(\begin{array}{cc} 
         M_2                   &  m_W\sqrt{2}\sin{\beta}\\ 
         m_W\sqrt{2}\cos{\beta}& \mu 
  \end{array}\right). 
\label{eq:charmassmatrix}
\end{eqnarray} 
At tree level, the chargino system depends on the SU(2) gaugino mass
parameter $M_2$, the higgsino mass parameter $\mu$, and the ratio
$\tan\beta=v_2/v_1$ of the vacuum expectation values of the two
neutral CP-even Higgs fields. We parametrise the CP
violation by the physical phase $\varphi_\mu$ of $\mu=|\mu|
e^{i\varphi_\mu} $, taking by convention $M_2$ real and positive,
absorbing its possible phase by redefining the fields.

\medskip

By diagonalising the chargino matrix~\cite{Haber:1984rc},
\begin{equation} 
  U^{\ast} M_{\tilde \chi} V^{\dagger} = 
  \textsl{diag}(m_{\tilde{\chi}^\pm_{1}},m_{\tilde{\chi}^\pm_{2}}),
\label{diag}   
\end{equation} 
we obtain the chargino masses, $m_{\tilde{\chi}^\pm_{2}} \geq
m_{\tilde{\chi}^\pm_{1}} \geq 0$, as well as their couplings.  In
Appendix~\ref{sec:parametrisation}, we give the analytic expressions
for the two independent, unitary diagonalisation matrices $U$ and $V$.
We shall use them for a qualitative understanding of the
chargino mixing in the presence of a non-vanishing CP phase
$\varphi_\mu\neq 0$. 

\medskip

At the ILC, chargino production $e^+e^-\to\tilde\chi^+_i\tilde\chi^-
_j$ proceeds at tree level via $\tilde\nu_e$ exchange in the
$t$-channel, and $Z, \gamma,$ exchange in the $s$-channel, see the
Feynman diagrams in Fig.~\ref{fig:1}.  Note that the photon
exchange contribution vanishes for non-diagonal chargino
production, $i\neq j$. The relevant terms in the MSSM Lagrangian for
chargino production
are~\cite{MoortgatPick:1998sk,Haber:1984rc,Choi:1998ut}
\begin{eqnarray}
  \mathscr{L}_{\gamma{\tilde\chi}_j^+{\tilde\chi}_i^-}&=&
                -eA_\mu\overline{{\tilde\chi}_i^+}\gamma^\mu{\tilde\chi}_j^+
                \delta_{ij},\qquad e>0,\label{eq1}\\ 
  \mathscr{L}_{e{\tilde\nu}_e{\tilde\chi}_i^+}&=& 
                  -gV_{i1}^\ast\overline{{\tilde\chi}_i^+}^C  
                 P_L  e \, \tilde\nu_e^\ast 
                +h.c., \label{eq2}\\  
   \mathscr{L}_{Z^0{\tilde\chi}_i^+{\tilde\chi}_j^-}&=& 
                \frac{g}{\cos\theta_w}Z_\mu 
                \overline{{\tilde\chi}_j^+}\gamma^\mu\! 
                \left[O^{'L}_{ji}P_L + 
                O^{'R}_{ji}P_R\right]\!{\tilde\chi}_i^+, \qquad
\end{eqnarray}  
with $i,j =1,2$, and the projectors $P_{L,R}=(1\mp\gamma^5)/2$.
In Eq.~(\ref{eq1}) ``$e$'' refers to the electron coupling, in
  Eq.~(\ref{eq2}) it refers to the electron spinor field.  
Furthermore
\begin{eqnarray} 
  O^{'L}_{ij}&=&-V_{i1}V_{j1}^\ast -\frac{1}{2}V_{i2}V_{j2}^\ast 
               +\delta_{ij}\sin^2\theta_w, \label{OprimeL}\\  
  O^{'R}_{ij}&=& +
               U_{i1}U_{j1}^\ast -\frac{1}{2}U_{i2}U_{j2}^\ast 
               +\delta_{ij}\sin^2\theta_w,\label{OprimeR}
\end{eqnarray} 
with the weak mixing angle $\theta_w$, and the weak coupling constant
 $g=e /\sin\theta_w$. 
For diagonal chargino production,  $i = j$,
the $Z$-chargino couplings are real, see Eqs.~\eqref{OprimeL},
\eqref{OprimeR},  and  the production amplitude has no CP-violating terms at
tree level.  

\medskip

For the subsequent chargino decay into the tau,
 ${\tilde\chi}_i^\pm\to \tau^\pm {\tilde\nu}_\tau^{(\ast)}$,
the contribution to the Lagrangian is~\cite{Haber:1984rc}
\begin{equation}
  \mathscr{L}_{{\tilde\chi}_i^+ \tau{\tilde\nu}_\tau}= 
                 -g\overline{{\tilde\chi}_i^+}^C
                  ( c^L_{i\tau} P_L  + c_{i\tau}^R P_R)
                  \tau{\tilde\nu}_\tau^\ast + h.c.,  
\label{taulagrangian}
\end{equation}
with the left and right $\tau$-$\tilde\nu_\tau$-chargino couplings
\begin{equation} 
  c^L_{i\tau} = V_{i1}^\ast, \qquad c^R_{i\tau} = - Y_\tau U_{i2},
\label{LRcouplings}
\end{equation}
and the Yukawa coupling
\begin{equation}
  Y_\tau     =\frac{m_\tau}{\sqrt{2}m_W\cos\beta}\,.
\label{Yukawa} 
\end{equation}
In the following, we present the analytical formulae for
the normal tau polarisation, which will be a sensitive probe of
CP violation in the chargino system.

\medskip

\subsection{Tau spin density matrix}\label{sub:matrix}

The unnormalised, $2\times 2$, hermitean, $\tau$ spin density matrix 
for chargino production, Eq.~\eqref{charprod}, and decay, 
Eq.~\eqref{chardec}, reads 
\begin{equation} 
  \rho^{\lambda_\tau\lambda^\prime_\tau}\equiv  
  \int\left(|\mathcal{M}|^2\right)^{\lambda_\tau\lambda^\prime_\tau}{\dLips}, 
\label{rho} 
\end{equation} 
with the amplitude $\mathcal{M}$, and the Lorentz invariant phase
space element ${\dLips}$, for details see
Appendix~\ref{sec:phasespace}.  The $\tau$ helicities are denoted by
$\lambda_\tau$ and $\lambda^\prime_\tau$. In the spin density matrix
formalism~\cite{Haber:1994pe}, the amplitude squared for the
  full process: production and decay, is given by\footnote{In
    the following, in order to avoid cluttering the notation, we drop
    the $\pm$ superscripts on the chargino in formulae: $\tilde\chi_i$.}
\begin{equation} 
  \left(|\mathcal{M}|^2\right)^{\lambda_\tau\lambda^\prime_\tau}= 
  |\Delta({\tilde{\chi}}_i)|^2\sum_{\lambda_i,\lambda^\prime_i} 
  (\rho_{\rm P})^{\lambda_i\lambda^\prime_i} 
  (\rho_{\rm D})_{\lambda^\prime_i\lambda_i}^{\lambda_\tau\lambda^\prime_\tau},
\label{densitymatrix} 
\end{equation} 
with the chargino helicities denoted $\lambda_i,\,
\lambda_i^\prime$, and an implicit sum over the helicities
of chargino ${\tilde\chi}_j$, whose decay is not further considered.
The amplitude squared decomposes into the remnant of the chargino
propagator,
\begin{equation} 
  \Delta(\tilde\chi_i)=\frac{1}{p_{\tilde\chi_i}^2-m_{\tilde\chi_i}^2 
  +i m_{\tilde\chi_i}\Gamma_{\tilde\chi_i}},
\label{propagator} 
\end{equation}
with mass $m_{{\tilde\chi}_i}$ and width $\Gamma_{{\tilde\chi}_i}$ of the
decaying chargino,  and the unnormalised spin density matrices 
$\rho_{\rm P}$ for production~(${\rm P}$), and $\rho_{\rm D}$ for 
 decay~(${\rm D}$). They can be expanded in terms of the Pauli
matrices $\sigma^a$
\begin{eqnarray} 
  (\rho_{\rm P})^{\lambda_i\lambda^\prime_i}&=&  
               2\left[{\rm P}\delta^{\lambda_i\lambda^\prime_i} 
              +\Sigma_{\rm P}^a(\sigma^a)^{\lambda_i\lambda^\prime_i}\right],
   \label{Pterm}\\ 
   (\rho_{\rm D})_{\lambda^\prime_i\lambda_i}^{\lambda_\tau\lambda^\prime_\tau}
    &=&   
               {\rm D}^{\lambda_\tau\lambda^\prime_\tau} 
                              \delta_{\lambda^\prime_i\lambda_i} 
               + {(\Sigma_{\rm D}^a)}^{\lambda_\tau\lambda^\prime_\tau} 
                              (\sigma^a)_{\lambda^\prime_i\lambda_i}, 
    \label{dms} 
\end{eqnarray} 
with an implicit sum over $a=1,2,3$.
For chargino production, the analytical formulae for the coefficients 
${\rm P}$  and $\Sigma_{\rm P}^a= \Sigma_{\rm P}^\mu s_{\tilde\chi_i,\mu}^a$,
which depend on the  chargino spin vectors $s_{\tilde\chi_i}^a$, are
explicitly given in Ref.~\cite{MoortgatPick:1998sk}. In that convention, a sum
over the helicities $\lambda_j,\lambda_j^\prime$ of chargino  
${\tilde\chi}^\mp_j$, whose decay we do not further consider, gives the factor
of~$2$  in Eq.~\eqref{Pterm}.  

\medskip

For the chargino ${\tilde\chi}^\pm_i$ decay into a tau, Eq.~\eqref{chardec},
we define  a set of tau spin basis vectors $s_\tau^b$, see
Appendix~\ref{sec:momenta}.  We then expand the coefficients of the decay 
matrix $\rho_{\rm D}$,  Eq.~\eqref{dms},
\begin{eqnarray} 
  {\rm D}^{\lambda_\tau\lambda^\prime_\tau}&=&  
               {\rm D}\delta^{\lambda_\tau\lambda^\prime_\tau} 
              +{\rm D}^b(\sigma^b)^{\lambda_\tau\lambda^\prime_\tau},\\ 
  (\Sigma_{\rm D}^a)^{\lambda^\prime_\tau\lambda_\tau}&=& 
               \Sigma_{\rm D}^a 
                              \delta^{\lambda_\tau\lambda_\tau^\prime} 
              +\Sigma_{\rm D}^{ab} 
                              (\sigma^b)^{\lambda_\tau\lambda_\tau^\prime}, 
\end{eqnarray} 
with an implicit sum over $b=1,2,3$.
A calculation of the expansion coefficients yields 
\begin{eqnarray} 
  {\rm D} &=& 
            \frac{g^2}{2}\left(|V_{i1}|^2+Y_\tau^2|U_{i2}|^2\right) 
             (p_{{\tilde\chi}_i}\cdot p_{\tau})\nonumber\\[2mm] 
           &&
             -g^2 Y_\tau \Re\{V_{i1}\ U_{i2}\}
             m_{\tilde\chi_i}m_\tau,\label{DD}\\[2mm] 
  {\rm D}^b &=& 
             \,^{\;\,+}_{(-)}\frac{g^2}{2} 
             \left(|V_{i1}|^2 - Y_\tau^2|U_{i2}|^2\right) 
             m_\tau(p_{{\tilde\chi}_i}\cdot s_{\tau}^b),\label{DDb}\\ 
   \Sigma_{\rm D}^a &=&
              \,^{\;\,-}_{(+)}\frac{g^2}{2} 
              \left(|V_{i1}|^2 - Y_\tau^2|U_{i2}|^2 \right) 
              m_{\tilde\chi_i}(p_{\tau}\cdot s_{{\tilde\chi}_i}^a),
              \label{SigmaD}\\[2mm] 
   \Sigma_{\rm D}^{ab} &=&
              g^2Y_\tau \Re\{ V_{i1}\ U_{i2}\}\times \nonumber\\[2mm]
                       &&
              \Big[ (p_{\tilde\chi_i}\cdot p_\tau)  
                    (s_{\tilde\chi_i}^a\cdot s_\tau^b)
                   -(p_\tau\cdot s_{\tilde\chi_i}^a) 
                    (p_{\tilde\chi_i}\cdot s_\tau^b)\Big]~\nonumber\\
                       && 
              - \frac{g^2}{2} \left(|V_{i1}|^2+Y_\tau^2|U_{i2}|^2\right)
                m_{\tilde\chi_i} m_\tau( s_{\tilde\chi_i}^a\cdot s_\tau^b)
                \nonumber\\[2mm] 
                        && 
                \,^{\;\,+}_{(-)} g^2Y_\tau \Im\{V_{i1}\ U_{i2}\} 
                \left[p_{\tilde\chi_i},\, p_\tau, \, 
                  s_{\tilde\chi_i}^a,\, s_{\tau}^b \right],
   \label{mycoeffs} 
\end{eqnarray} 
with the weak coupling constant $g$, and the Yukawa coupling $Y_\tau$,
\textit{cf.} Eq.~\eqref{Yukawa}.  The formulae are given for
the decay of a positive chargino, $\tilde\chi_i^+\to\tau^+\tilde\nu_
\tau$. The signs in parentheses hold for the charge conjugated decay
$\tilde\chi_i^- \to \tau^- \tilde\nu^\ast_\tau$.

\medskip

The spin-spin correlation term $\Sigma^{ab}_{\rm D}$ in
Eq.~\eqref{mycoeffs} explicitly depends on the imaginary part
$\Im\{V_{i1}\ U_{i2}\} $ of the chargino matrices $U$ and $V$,
\textit{cf.} Eqs.~\eqref{U} and \eqref{V} in
Appendix~\ref{sec:parametrisation}.   
Thus this term is manifestly CP-sensitive, \textit{i.e.}
sensitive to the phase $\varphi_\mu$ of the chargino sector. The
imaginary part is multiplied by the totally anti-symmetric (epsilon)
product
\begin{equation}
  \mathcal{E}^{ab}\equiv
  \left[p_{\tilde\chi_i},\, p_\tau, \, s_{\tilde\chi_i}^a,\, s_{\tau}^b \right]
                  \equiv 
  \epsilon_{\mu\nu\rho\sigma} \, p_{\tilde\chi_i}^\mu \,
           p_{\tau}^\nu \, s_{\tilde\chi_i}^{a,\rho} \, s_{\tau}^{b,\sigma}.  
  \label{eq:epsilon}  
\end{equation}
We employ the convention for the epsilon tensor $\epsilon_{0123}=1$.
Since each of the spacial components of the four-momenta $p$,
or spin vectors $s$, changes sign under a naive time transformation, $t \to
-t$, the  epsilon product $\mathcal{E}^{ab}$ is T-odd.\\

\medskip

Inserting the density matrices, Eqs.~\eqref{Pterm} and \eqref{dms}, 
into Eq.~\eqref{densitymatrix}, we get for the amplitude squared
\begin{eqnarray} 
(|\mathcal{M}|^2)^{\lambda_\tau\lambda_\tau^\prime}&=& 
                     4|\Delta({\tilde\chi}_i)|^2\times\nonumber\\
&&                 \Big[~({\rm PD} + \Sigma_{\rm P}^a\Sigma_{\rm D}^a) 
                          \delta^{\lambda_\tau\lambda_\tau^\prime} \nonumber\\ 
&&                      +({\rm PD}^b + \Sigma_{\rm P}^a\Sigma_{\rm D}^{ab})   
                         (\sigma^b)^{\lambda_\tau\lambda_\tau^\prime}\Big],
\label{nosum} 
\end{eqnarray}
with an implicit sum over $a,b=1,2,3$.
Note that the terms proportional to $m_\tau$ in
Eqs.~\eqref{DD}, \eqref{DDb}, and \eqref{mycoeffs}, are negligible at high 
particle energies $E \gg m_\tau$, in particular  ${\rm D}^b$ 
can be neglected in Eq.~\eqref{nosum}.

\medskip

\subsection{Normal tau polarisation and CP asymmetry}
                                                    \label{sub:asymmetry}
The $\tau$ polarisation for the overall event sample 
is given by the expectation value of the Pauli matrices
{\boldmath $\sigma$}$=(\sigma^1,\sigma^2,\sigma^3)$~\cite{Renard:1981de}

\begin{equation} 
  \mbox{\boldmath $ \mathcal P $ }
  = \frac{{\rm Tr}\{\rho \mbox{\boldmath$ \sigma$} \}}{{\rm Tr}\{\rho\}}, 
\label{taupol} 
\end{equation} 
with the $\tau$ spin density matrix $\rho$, as given in  Eq.~\eqref{rho}.
In our convention for the polarisation vector
{\boldmath $ \mathcal P $ }$ =(\mathcal{P}_1,\mathcal{P}_2,\mathcal{P}_3)$, 
the components $\mathcal{P}_1$ and $\mathcal{P}_3$ are the transverse and
longitudinal  
polarisations in the plane spanned by ${\bf p}_{\tau}$, and ${\bf p}_{e^-}$, 
respectively, and  $\mathcal{P}_2$  is the polarisation normal to that plane.
See our definition of the tau spin basis vectors $s_\tau^b$ in
Appendix~\ref{sec:momenta}. 

\medskip

The normal $\tau$ polarisation is equivalently defined as
\begin{eqnarray} 
  \mathcal{P}_2 &\equiv&
  \frac{N(\uparrow)-N(\downarrow)}{N(\uparrow)+N(\downarrow)},
\label{tau_norm_pol}
\end{eqnarray}
with the number of events $N$ with the 
$\tau$ spin up $(\uparrow)$ or down $(\downarrow)$, with respect
to the quantisation axis  ${\bf p}_{\tau}\times{\bf p}_{e^-}$,
\textit{cf.}  Eq.~\eqref{tauspin} in Appendix~\ref{sec:momenta}.
The normal  $\tau$ polarisation can  thus also be regarded as an asymmetry
\begin{equation} 
  \mathcal{P}_2\equiv\mathcal{A}^{\rm T}=
                  \frac{\sigma(\mathcal{T}>0)-\sigma(\mathcal{T}<0)} 
                   {\sigma(\mathcal{T}>0)+\sigma(\mathcal{T}<0)}\,, 
\label{tripleasymm}
\end{equation} 
of the triple product 
\begin{equation} 
  \mathcal{T}={\mbox{\boldmath$ \xi$} }_{\tau}\cdot
              ({\bf p}_{\tau}\times{\bf p}_{e^-})\,, 
\label{triple} 
\end{equation} 
where {\boldmath$ \xi$}$_{\tau}$ is the direction of the $\tau$ spin
vector for each event. 
The triple product $\mathcal{T}$ is included in
the spin-spin correlation term $\Sigma_{\rm P}^a \Sigma_{\rm D}^{ab}$
of the amplitude squared $|\mathcal{M}|^2$, Eq.~\eqref{nosum}. 
The asymmetry thus probes the term which contains the epsilon
product $\mathcal{E}^{ab}$, Eq.~\eqref{eq:epsilon}.

\medskip

Since under naive time reversal, $t\to -t$, the triple product
$\mathcal{T}$ changes sign, the tau polarisation $\mathcal{P}_2$,
Eq.~(\ref{tripleasymm}), is T-odd.  Due to CPT
invariance~\cite{Luders:1954zz}, $\mathcal{P}_2$ would thus be CP-odd
at tree level.  In general, $\mathcal{P}_2$ also has contributions
from absorptive phases, \textit{e.g.} from intermediate
$s$-state resonances or final-state interactions, which do not signal
CP violation. Although such absorptive contributions are a higher
order effect, and thus expected to be small, they can be eliminated in
the true CP asymmetry~\cite{Bartl:2003gr}
\begin{equation} 
  \mathcal{A}_\tau^{\rm CP} = \frac{1}{2}(\mathcal{P}_2 - 
\overline{\mathcal{P}}_2), 
\label{asym} 
\end{equation} 
where $\overline{\mathcal{P}}_2$ is the normal tau polarisation for the
charged  
conjugated process ${\tilde{\chi}}_i^-\to\tau^-{\tilde{\nu}}_\tau^\ast$.  
For our analysis at tree level,  where no absorptive phases are present,
we have $\mathcal{A}_\tau^{\rm CP}=\mathcal{P}_2$.
Thus we will study $\mathcal{A}_\tau^{\rm CP}$ in the following, which
is however equivalent to  $\mathcal{P}_2$ at tree level.

\medskip

Inserting now the explicit form of the density matrix $\rho$, Eq.~\eqref{rho},
into Eq.~\eqref{taupol}, together with Eq.~\eqref{nosum}, we obtain  the
CP asymmetry     
\begin{equation} 
\mathcal{A}_\tau^{\rm CP}=\mathcal{P}_2 
                         =\frac{\int \Sigma_{\rm P}^a\Sigma_{\rm
                         D}^{ab=2}\,{\dLips}} 
                               {\int  {\rm PD}\,{\dLips}},  
\label{P2} 
\end{equation} 
where we have used the narrow width approximation for the propagators
in the phase space element ${\dLips}$, see Eq.~\eqref{nwapprox}. 
Note that in the denominator 
of $\mathcal{A}_\tau^{\rm CP}$, Eq.~\eqref{P2}, all spin correlation terms 
vanish, $\int  \Sigma_{\rm P}^a\Sigma_{\rm D}^a  \,{\dLips}=0$,
when integrated over phase space. In the numerator  only the spin-spin
correlation term  $\Sigma_{\rm P}^a \Sigma_{\rm D}^{ab=2}$ contributes,
since only $\Sigma_{\rm D}^{ab}$, Eq.~\eqref{mycoeffs}, contains the T-odd 
epsilon\footnote{Note that for $\tilde\chi_i^+\tilde\chi_j^-$ production, 
      with  $i\neq j$, there is, in principle, also a contribution from the
      CP-violating normal chargino polarisation $\Sigma_{\rm P}^{a=2}$ to the
      asymmetry 
      $\mathcal{A}_\tau^{\rm CP}$, which projects out the CP-even parts of 
      $\Sigma_{\rm D}^{ab=2}$.  However, $\Sigma_{\rm P}^{a=2}$ is
      numerically small in our scenarios with large $\tan\beta$,
      so that we do not discuss its impact here; see
      Refs.~\cite{Kittel:2004rp,CHAR2} for
      CP asymmetries in chargino production 
       } 
product $\mathcal{E}^{ab}$, see Eq.~\eqref{eq:epsilon}.

\subsection{Parameter dependence of the CP asymmetry}
\label{sub:analytics}

To qualitatively understand the dependence of the asymmetry
$\mathcal{A}_\tau^{\rm CP}$, Eq.~\eqref{P2}, on the parameters of the
chargino sector, we study in some detail its dependence on the
$\tau$-$\tilde\nu_\tau$-chargino couplings, $c^L_{i\tau}$ and
$c^R_{i\tau}$, \textit{cf.}  Eq.~\eqref{taulagrangian}.  From
the explicit form of the decay terms ${\rm D}$, Eq.~\eqref{DD}, and
$\Sigma_{\rm D}^{ab}$, Eqs.~\eqref{mycoeffs}, \eqref{eq:epsilon}, we
find that the asymmetry

\begin{equation}
  \mathcal{A}_\tau^{\rm CP}=\eta_i
           \frac{\int \Sigma_{\rm P}^a
                \; \mathcal{E}^{ab=2}
              \,\dLips}
              {\int {\rm P} \, (p_{{\tilde\chi}_i}\cdot p_{\tau}) \,\dLips}\,,
\label{finalmathcalA}
\end{equation}
with $(p_{{\tilde\chi}_i}\cdot p_{\tau})=(m_{{\tilde\chi}_i}^2-m_
{\tilde\nu_\tau}^2)/2$, is proportional to the decay coupling factor 
\begin{equation}
\eta_i=
\frac{  \Im\{c^L_{i\tau} c^{R \ast}_{i\tau} \} }
          {\frac{1}{2}(|c^L_{i\tau}|^2 + |c^R_{i\tau} |^2)}
=
\frac{Y_\tau\Im\{V_{i1} U_{i2}\}}
          {\frac{1}{2}(|V_{i1}|^2+Y_\tau^2 |U_{i2}|^2)}\,,
\label{etafactor}
\end{equation}
with $\eta_i \in [-1,1]$.
Using the explicit forms of the matrix elements $U$ and $V$, see
Eqs.~\eqref{U} and~\eqref{V}, we can 
rewrite the factor $\eta_i$ 
for  ${\tilde\chi}_1^\pm \to\tau^\pm{\tilde\nu}_\tau^{(\ast)}$,
or ${\tilde\chi}_2^\pm \to \tau^\pm{\tilde\nu}_\tau^{(\ast)}$
decay, respectively
\begin{eqnarray}
  \eta_1 &=&
  \frac{- Y_\tau\sin(\theta_1) \cos(\theta_2)} 
       {\frac{1}{2}\left[\cos^2(\theta_2)+ Y_\tau^2 \sin^2(\theta_1)\right]} 
                   \, \sin(\gamma_1+\phi_1)\,,
\label{eta1}\\
\eta_2 &=&
\frac{Y_\tau \cos(\theta_1) \sin(\theta_2) }
                  {\frac{1}{2}[\sin^2(\theta_2)+Y_\tau^2 \cos^2(\theta_1)]} 
                      \, \sin(\gamma_2+\phi_2)\,,
\label{eta2}
\end{eqnarray}
with the angles $\theta_{1,2}$, which describe chargino mixing, and
$\gamma_{1,2}$ and $\phi_{1,2}$ which describe their CP properties,
\textit{cf.} Eqs.~\eqref{uvangles1} and~\eqref{uvangles2}
in App.~\ref{sec:parametrisation}.

\medskip

\begin{figure*}[h!]  
\subfigure[]{  
  \includegraphics[width=0.94\columnwidth]{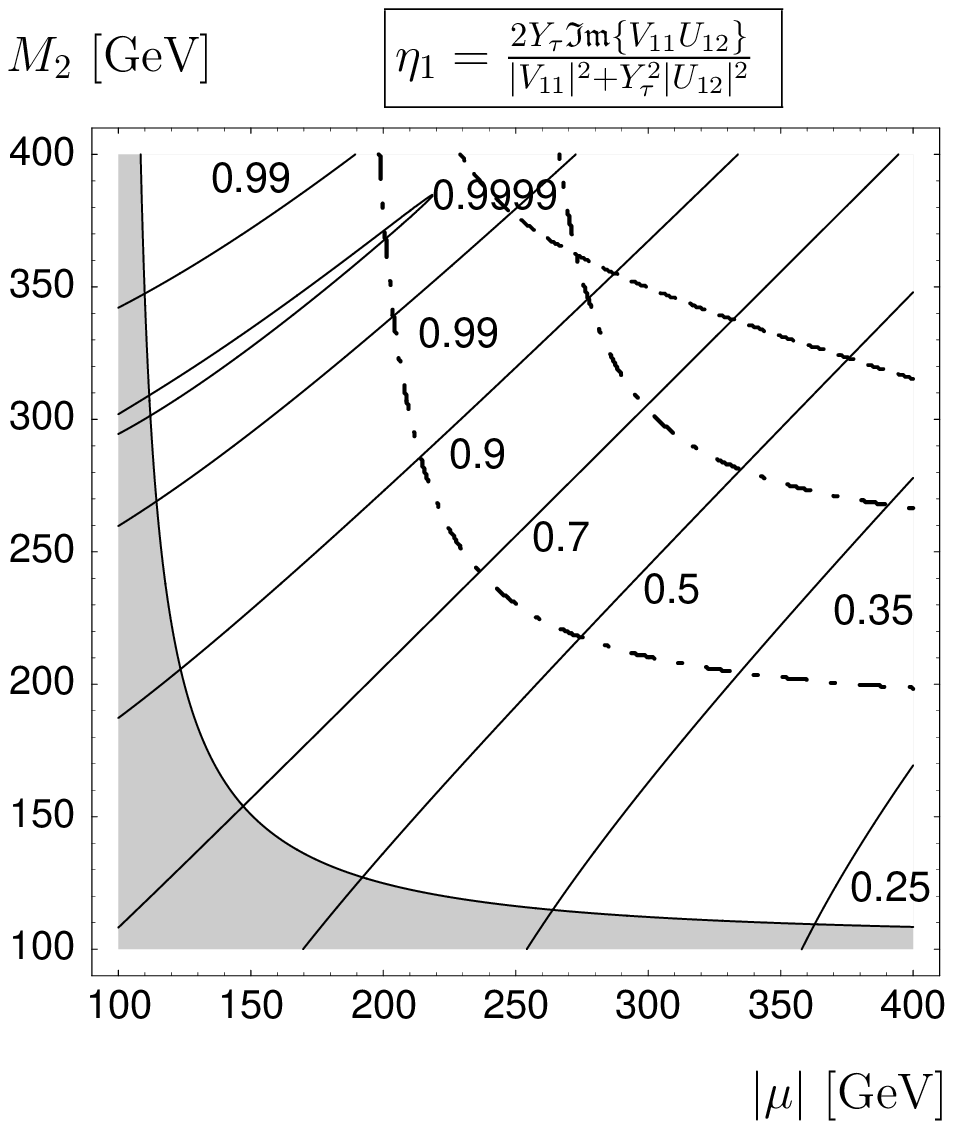}}  
\subfigure[]{  
  \includegraphics[width=0.94\columnwidth]{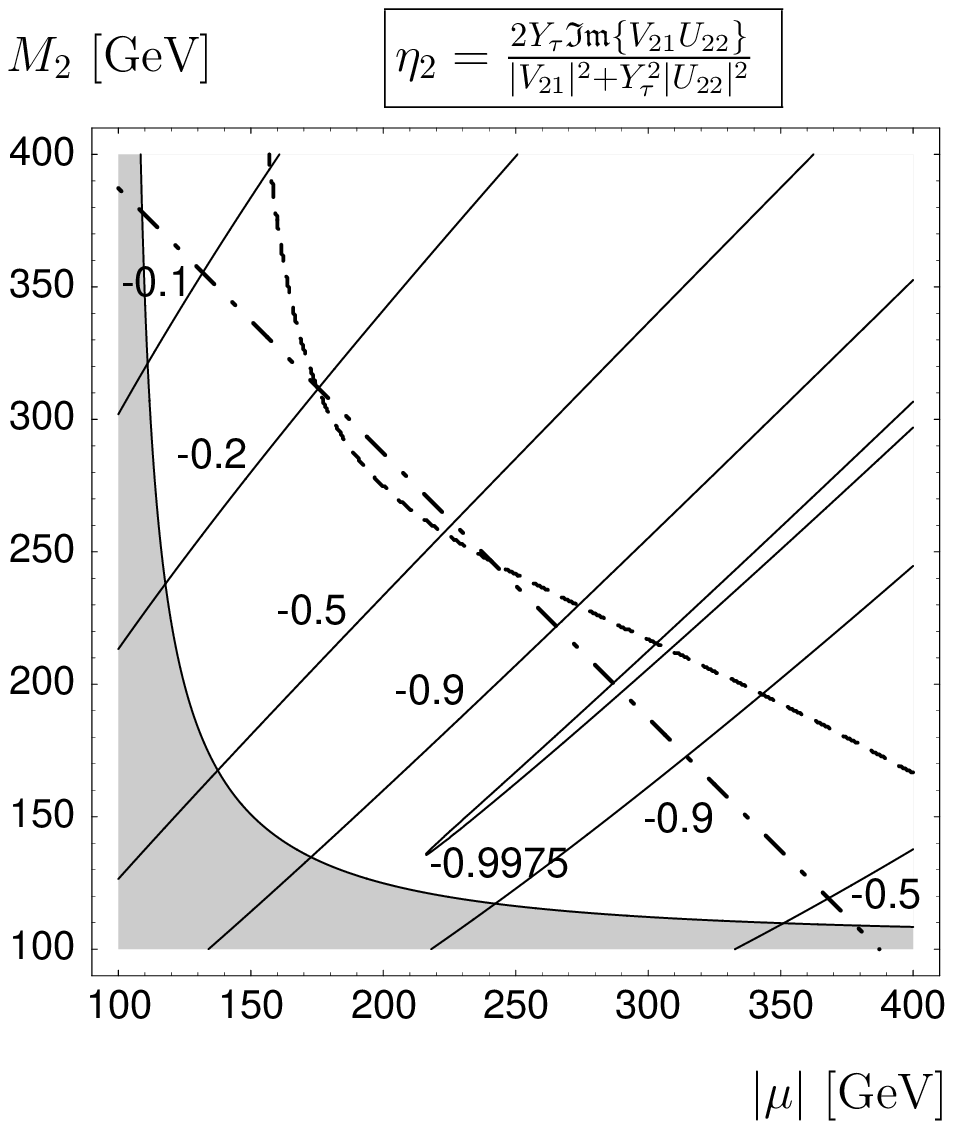}}  
\caption{Contour lines in the $M_2$--$|\mu|$ plane for (a) the
  proportionality factor $\eta_1$, Eq.~\eqref{eta1}, of the asymmetry
  $\mathcal{A}_\tau^{\rm CP}$ for $e^+ e^- \to {\tilde\chi}_1^+
  {\tilde\chi}_1^-$, ${\tilde\chi}_1^\pm \to \tau^\pm
  {\tilde\nu}_\tau^{(\ast)}$ and the scenario as defined in
  Table~\ref{tab:1}, (b) the proportionality factor $\eta_2$,
  Eq.~\eqref{eta2}, of the asymmetry $\mathcal{A}_\tau^{\rm CP}$ for
  $e^+ e^- \to {\tilde\chi}_2^+ {\tilde\chi}_1^-$, ${\tilde\chi}_
  \tau^+ {\tilde\nu}_\tau$ and the scenario as given in
  Table~\ref{tab:1}. In each case with a centre-of-mass energy
  $\sqrt{s}=500$~GeV.  Above the dashed line the lightest neutralino
  is no longer the LSP since
  $m_{{\tilde\tau}_1}<m_{{\tilde\chi}_1^0}$.  In the grey-shaded area
  $m_{{\tilde\chi}_1^\pm}<104$~GeV.  In (a), the band between the
  dashed-dotted lines, and in (b) the triangle below the dashed-dotted
  line mark the kinematically allowed regions, see
  Figs.~\ref{fig:3}(c) and \ref{fig:6}(c), respectively.}
\label{fig:2}  
\end{figure*}  

Since $\eta_i$ is  proportional to the Yukawa coupling $Y_\tau$, 
Eq.~\eqref{Yukawa}, the asymmetry will be enhanced for increasing $\tan\beta$.
Then the phase dependence of the asymmetries will be
$\mathcal{A}_\tau^{\rm CP}\propto \eta_i \propto \sin(\gamma_i+\phi_i)\approx 
\sin(\varphi_\mu)$,
since we find  $\phi_1,\gamma_2 \to \varphi_\mu$ and   $\phi_2,\gamma_1\to 0$ 
for $\tan\beta\gg 1$.
Note that the asymmetry will be additionally suppressed if $\tan\beta$ is
small, since  that not only results in $Y_\tau\ll 1$, but also
leads to $\phi_1,\gamma_1\to 0$, and $\phi_2\to -\varphi_\mu$, 
 $\gamma_2\to\varphi_\mu$, which means that the phase dependent part
of $\eta_i$ vanishes,  $\sin(\gamma_i+\phi_i)\to 0$.

\medskip

Second, we expect maximal asymmetries for equal 
left and right chargino couplings, $|c^L_{i\tau}| \approx |c^R_{i\tau}|$,
where the coupling factor can be  maximal $\eta_i=\pm1$, see
Eq.~\eqref{etafactor}.
Concerning the mixing of the charginos, parametrised  by the
angles $\theta_{1,2}$, we expect maximal asymmetries 
in a mixed gaugino-higgsino scenario, $|\mu|\approx M_2$, however 
``corrected'' by the Yukawa coupling,  such that 
$ \cos(\theta_2)\approx Y_\tau \sin(\theta_1)$ for ${\tilde\chi}_1^\pm$ decay,
and $ \sin(\theta_2)\approx Y_\tau \cos(\theta_1)$ for ${\tilde\chi}_2^\pm$
decay, see Eqs.~\eqref{eta1} and~\eqref{eta2}, respectively.

In Figs.~\ref{fig:2}(a) and (b), we show the $\eta$ factors $\eta_1$ and
$\eta_2$ in the $M_2$-$\mu$ plane for the scenarios as given in
Table~\ref{tab:1} and \ref{tab:2}, accordingly. 

\section{Numerical results}\label{sec:results}

We present numerical results for the CP asymmetry $\mathcal{A}_\tau^
{\rm CP}$, Eq.~(\ref{P2}), for chargino production $e^+e^-\to{\tilde
  \chi}_i^\pm{\tilde\chi}_j^\mp$ and decay ${\tilde{\chi}}_i^\pm\to
\tau^\pm {\tilde{\nu}}^{(\ast)}_\tau$.  We choose a centre-of-mass
energy of $\sqrt{s}=500$~GeV and longitudinally polarised beams
$(\mathcal{P}_{e^-}|\mathcal{P}_{e^+})= (-0.8|0.6)$.
We include all spin correlations between
chargino production and decay, since only they include CP-violating
terms at tree level.

\medskip
  
We study the dependence of the CP asymmetry on the MSSM parameters  
$\mu=|\mu|e^{i\varphi_\mu}$, $M_2$, and $\tan\beta$. For 
${\tilde\chi}_1^-{\tilde\chi}_1^+$ production, we study  the dependence on the 
beam polarisations.  The feasibility of measuring the asymmetry also  depends 
on the chargino production  cross sections and decay branching ratios, which 
we discuss in detail. Finally, to get a lower  bound on the event rates 
necessary to observe the CP asymmetry, we also give  its theoretical 
statistical significance $S_\tau$, Eq.~\eqref{significance}.\\   
  
For the calculation of the chargino decay widths and branching ratios, we  
consider the two-body decays~\cite{Kittel:2004rp}  
\begin{eqnarray}  
{\tilde\chi}_i^+&\to&  
                     W^+{\tilde\chi}_k^0\,,\nonumber\\   
                &   &\tilde{e}_L^+\nu_e\,,\,\tilde{\mu}_L^+\nu_\mu\,,\,  
                     \tilde{\tau}_{1,2}^+\nu_\tau\,,\nonumber\\  
                &   &e^+{\tilde\nu}_e\,,\, \mu^+{\tilde\nu}_\mu\,,\,   
                     \tau^+{\tilde\nu}_\tau\,,\nonumber\\  
{\tilde\chi}_2^+&\to&{\tilde\chi}_1^+Z^0,\,{\tilde\chi}_1^+H_1^0\,,  
\label{chardecyas} 
\end{eqnarray}  
for $i=1,2$ and $k=1,\ldots,4$.  We neglect three-body decays, which
are suppressed by phase-space.  In order to reduce the number of
parameters, we use the GUT inspired relation $|M_1|=5/3\,M_2\tan^2
\theta_w$. This significantly constrains the neutralino
sector~\cite{Neutpaper}.  We take stau mixing into account, and set
the mass of the trilinear scalar coupling parameter to $A_\tau=250~
$GeV.  Since its phase does not contribute to the CP asymmetry, we fix
it to $\varphi_{A_\tau}=0$, as well as $\varphi_1=0$, which is the
phase of the gaugino mass parameter $M_1$.  We fix the soft-breaking
parameters $M_{\tilde E}=M_{\tilde L}$ in the slepton sector. When
varying $\mu$ and $M_2$, this can lead to excluded regions in the
plots where the lightest stau ${\tilde\tau}_1$ is the LSP, $m_{{
    \tilde\tau}_1}<m_{{\tilde\chi}_1^0}$, which we indicate by a
dashed line, \textit{cf.} Fig.~\ref{fig:2}. The Higgs mass
parameter is fixed to $M_A= 1000~$GeV. Our results are insensitive to this
choice, as long as we stay in the decoupling limit. 

\medskip

In order to show the full phase dependence on $\varphi_\mu$, we relax
the constraints from the electric dipole moments (EDMs).  Our purpose
is to demonstrate that a non-vanishing CP phase in the chargino sector
would lead to large asymmetries. Their measurement in chargino decays
will be a sensitive probe to constrain the phase $\varphi_\mu$
independently from EDM measurements.  There can be cancellations
between different contributions to the EDMs, which can in principle be
achieved by tuning parameters and CP phases
outside the chargino sector~\cite{cancellations1,cancellations2,cancellations3}.

\medskip

On the other hand, large phases can be in agreement with the EDM
bounds for scenarios with heavy first and second generation sfermion
masses, of the order of $10$~TeV. The third generation can stay light
with masses of the order of $100$~GeV~\cite{Bagger:1999ty}.
Such scenarios are particularly interesting also for our process. Heavy
electron sneutrinos enhance chargino production cross sections, while at the
same time the decay channels into taus, ${\tilde{\chi}}_i^\pm \to
\tau^\pm {\tilde{\nu}}^{(\ast)}_\tau$, will be dominating. We will
study such a scenario at the end of the numerical section.

\subsection{Chargino pair production $\boldsymbol{e^+ +  e^- \to  
            {\tilde\chi}_1^+ + {\tilde\chi}_1^-$ and decay  
            ${\tilde\chi}_1^+ \to \tau^+ + {\tilde\nu}_\tau}$}   
                                                  \label{sub:11prod1dec}
\begin{table}[b!]  
\caption{Scenario for chargino production $e^+ e^- \to {\tilde\chi}_1^+   
        {\tilde\chi}_1^-$ and decay ${\tilde\chi}_1^\pm \to \tau^\pm  
        {\tilde\nu}_\tau^{(\ast)}$. The mass parameters $M_2$, $|\mu|$,  
        $M_{\tilde E}$ and $M_{\tilde L}$ are given in GeV.}  
\begin{tabular}{cp{0.03\columnwidth}cp{0.05\columnwidth}cp{0.09\columnwidth}  
                cp{0.06\columnwidth}c}
\\  
\hline\hline  
${\tan\beta}_{\phantom{I}}$ && ${\varphi_\mu}_{\phantom{I}}$ && 
${M_2}_{\phantom{I}}$ && ${|\mu|}_{\phantom{I}}$  &&    
${M_{\tilde E}}_{\phantom{I}}=M_{\tilde L}$   
\\ 
\hline  
25 && $0.5\pi$ && 380 && 240 && 200   
\\  
\hline\hline  
&&&&&&&&  
\\  
\multicolumn{9}{c}{Calculated mass spectrum.}  
\\ 
\hline\hline  
${\tilde\ell}$                && $m$~[GeV]&&&&$\tilde\chi$    &&$m$~[GeV]\\ 
\hline    
${\tilde e}_R,{\tilde \mu}_R$  && 205     &&&&${\tilde\chi}_1^0$&& 175\\  
${\tilde e}_L,{\tilde \mu}_L$  && 206     &&&&${\tilde\chi}_2^0$&& 238\\  
${\tilde\nu}_e,{\tilde\nu}_\mu$&& 189     &&&&${\tilde\chi}_3^0$&& 247\\  
${\tilde\tau}_1$               && 177     &&&&${\tilde\chi}_4^0$&& 405\\  
${\tilde\tau}_2$               && 230     &&&&${\tilde\chi}_1^\pm$&& 225\\ 
${\tilde\nu}_\tau$             && 189     &&&&${\tilde\chi}_2^\pm$&& 405\\ 
\hline\hline  
\multicolumn{3}{c}{${\rm BR}({\tilde\chi}_1^+\to\tau^+\tilde\nu_\tau)$~[\%]\qquad}&&&    
\multicolumn{4}{c}{\enspace\enspace\enspace\enspace$\sigma_P(e^+e^-\to  
                   {\tilde\chi}_1^-{\tilde\chi}_1^+)$~[fb]}\\   
\hline  
                                &&\enspace 30&&&&                 &&417\\  
\hline\hline  
\end{tabular}  
\label{tab:1}  
\end{table}

We first study the production of the lightest chargino pair $e^+
e^-\to \tilde\chi_1^+\tilde\chi_1^-$. 
Note that the production amplitude 

\begin{figure*}[h!]  
\subfigure[]{  
  \includegraphics[width=0.97\columnwidth]{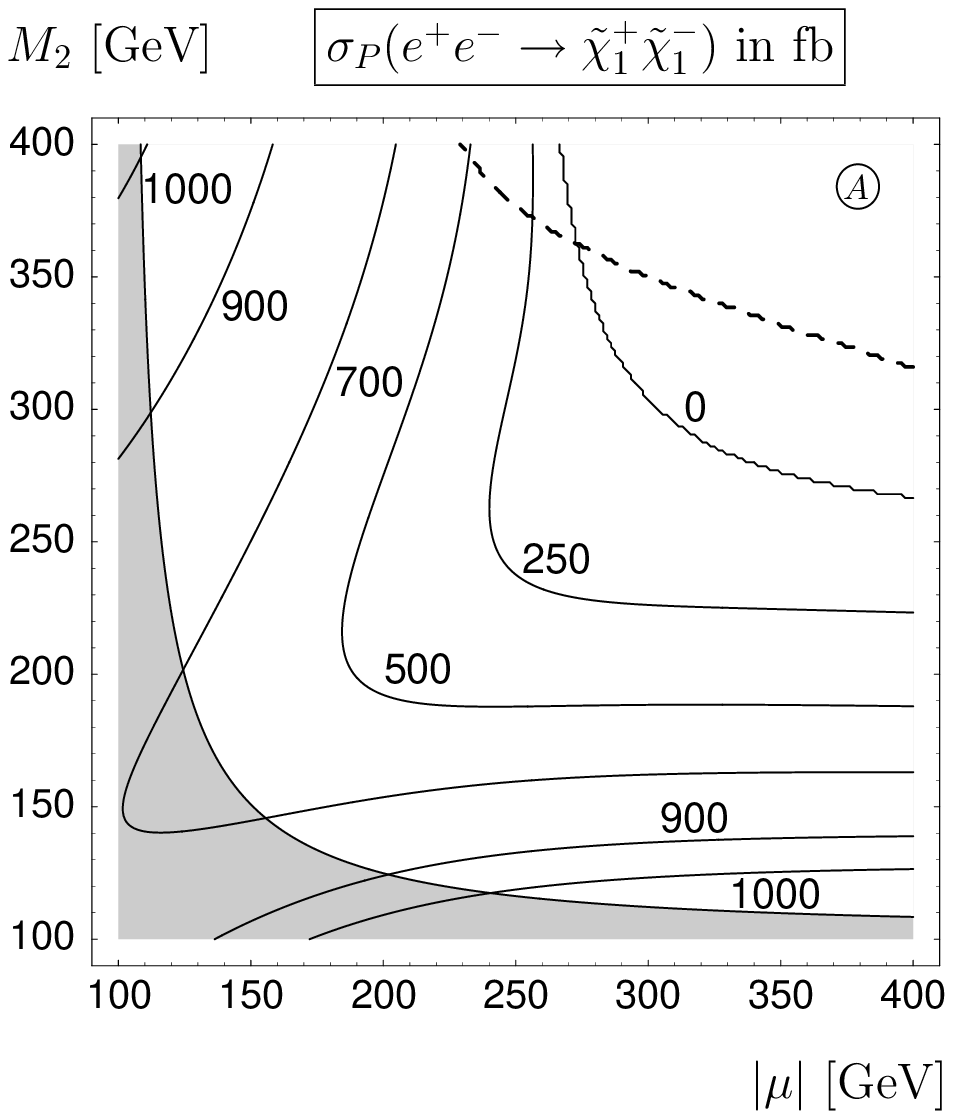}}  
  \hfill  
\subfigure[]{  
  \includegraphics[width=0.97\columnwidth]{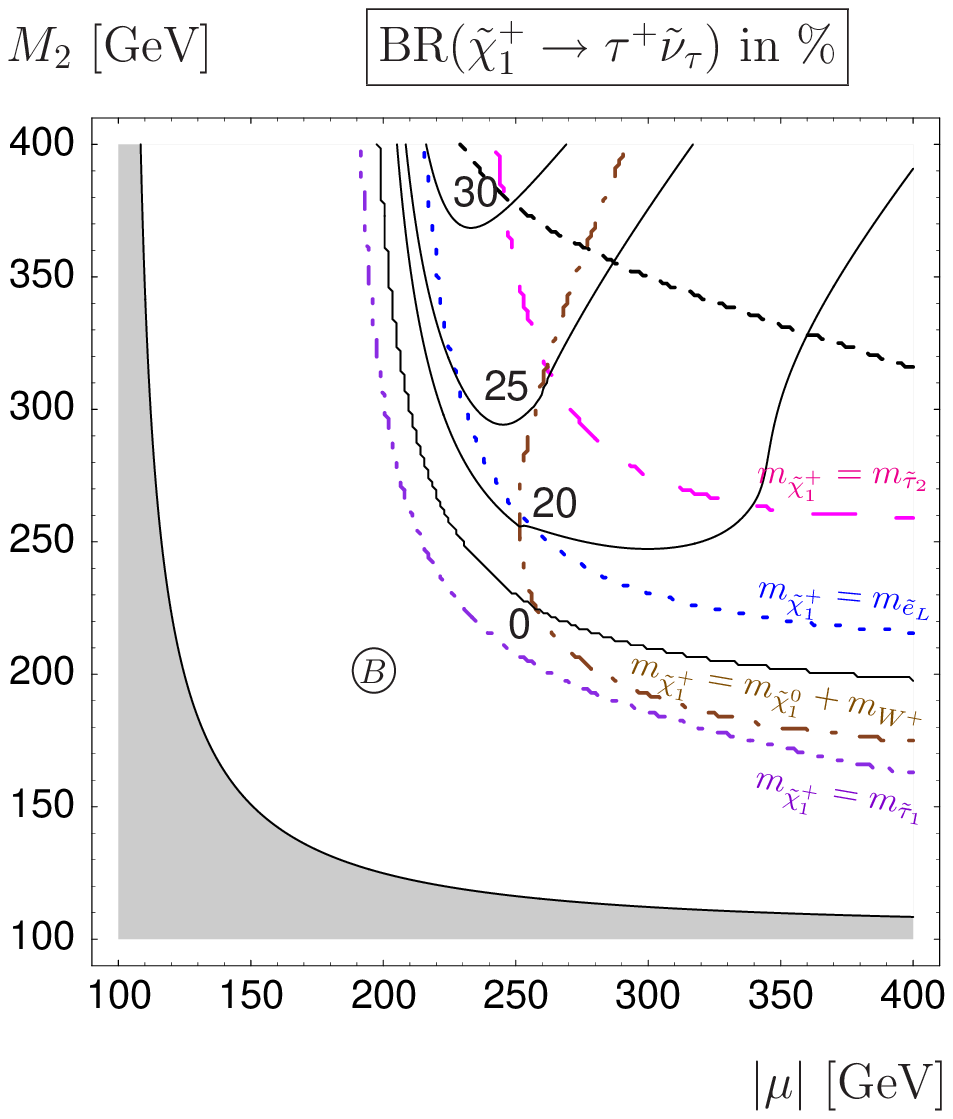}}  
\subfigure[]{  
  \includegraphics[width=0.97\columnwidth]{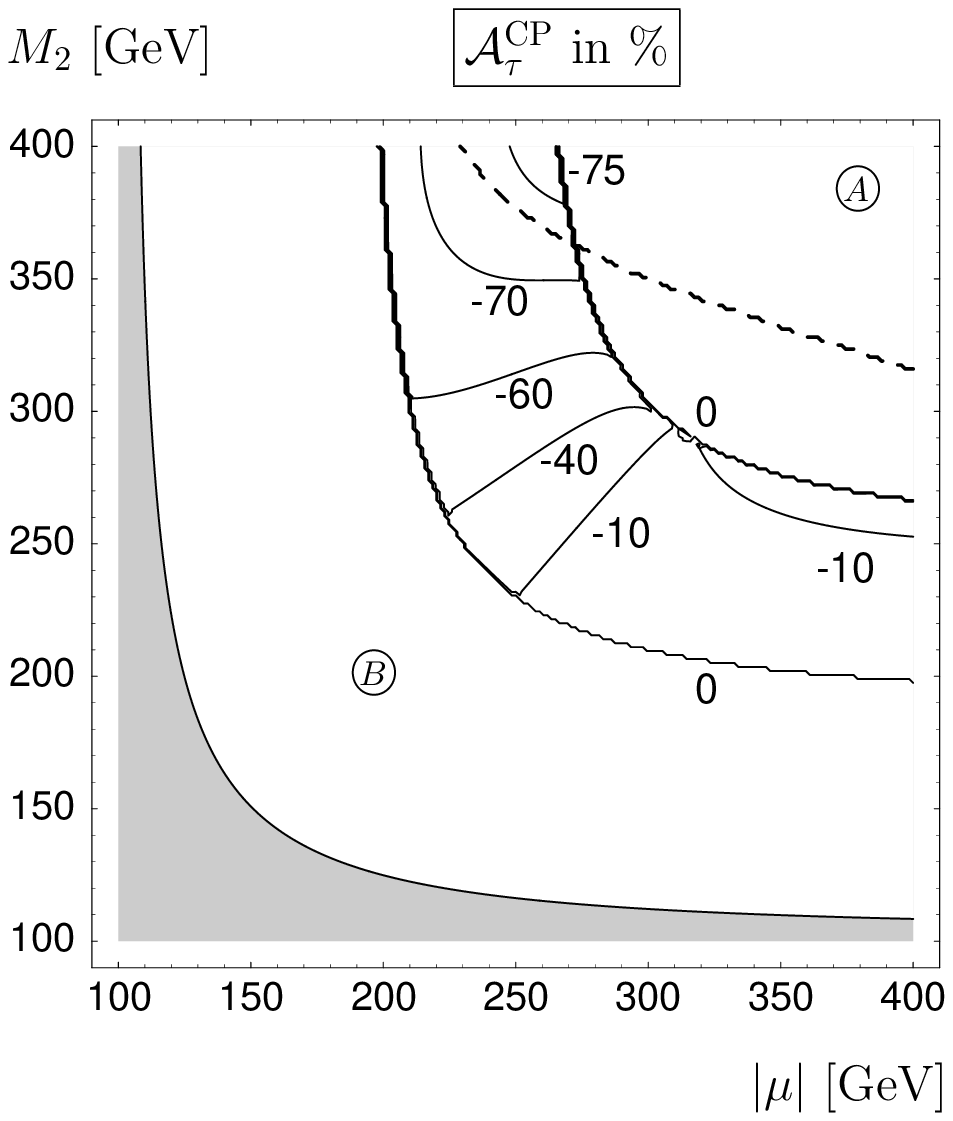}}  
  \hfill  
\subfigure[]{  
  \includegraphics[width=0.97\columnwidth]{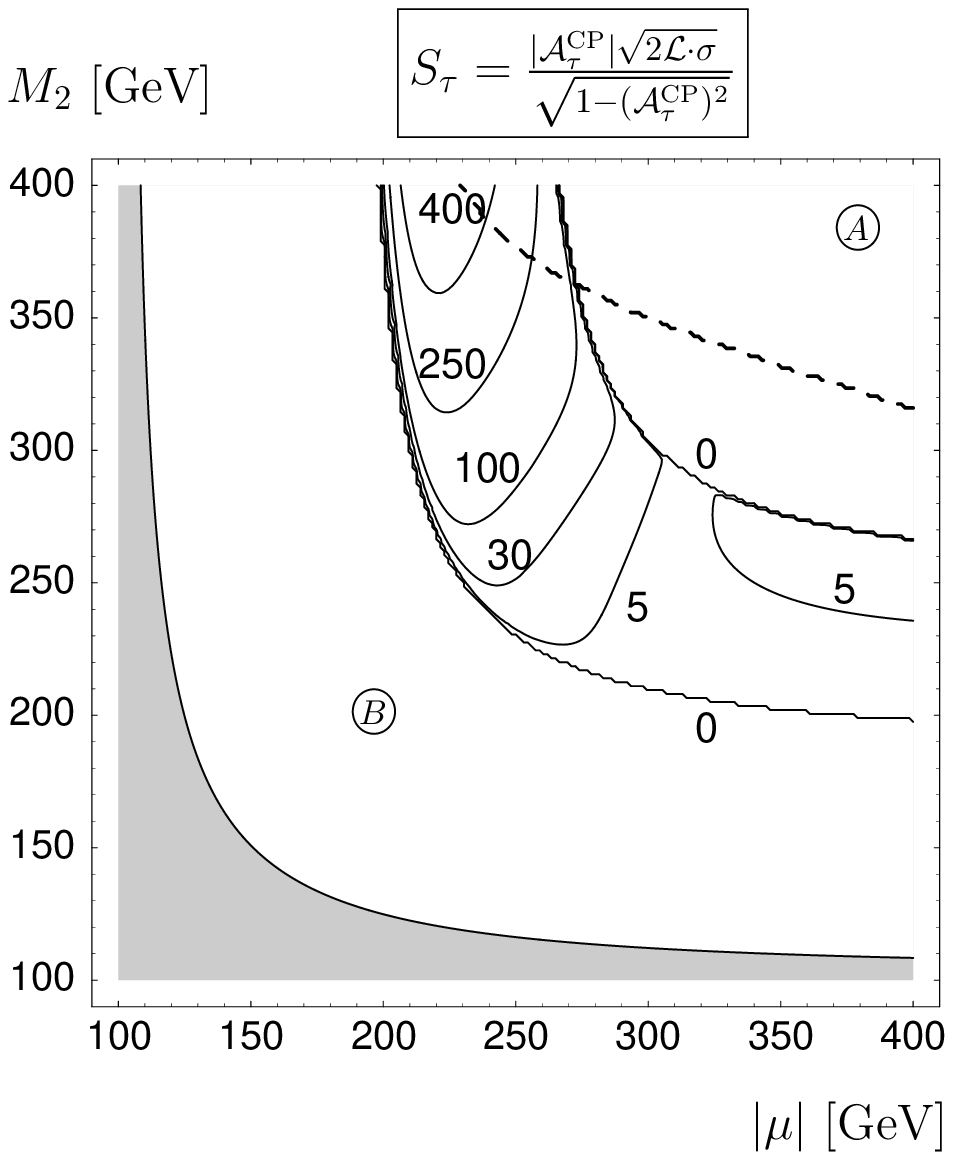}}  
\caption{Contour lines in the $M_2$--$|\mu|$ plane of (a) the production cross
         section, (b) branching ratio, (c) CP asymmetry of the normal tau
         polarisation, and (d) its significance for $e^+ e^- \to
         {\tilde\chi}_1^+ {\tilde\chi}_1^-$, ${\tilde\chi}_1^\pm \to   
         \tau^\pm {\tilde\nu}_\tau^{(\ast)}$,  with a  centre-of-mass energy
         $\sqrt{s}=500$~GeV, longitudinally polarised beams
         $(\mathcal{P}_{e^-}|\mathcal{P}_{e^+})=(-0.8|0.6)$, and an integrated
         luminosity $\mathcal{L}=500~{\rm fb}^{-1}$. The other SUSY parameters
         are defined in Table~\ref{tab:1}. The area
         {\large$\bigcirc$}\hspace{-0.36cm}{\small $A$}\,  
         above the zero contour line of the production cross section is  
         kinematically forbidden by $\sqrt{s}< 2 m_{{\tilde\chi}_1^\pm}$,  
          and the area {\large$\bigcirc$}\hspace{-0.355cm}{\small $B$}\,\,   
         below the  zero contour line of the branching ratio 
         by $m_{{\tilde\chi}_1^+}< m_{{\tilde\nu}_\tau}$. Above the  
         dashed line the lightest neutralino is no longer the LSP since  
         $m_{{\tilde\tau}_1}<m_{{\tilde\chi}_1^0}$. In   
         the grey-shaded area $m_{{\tilde\chi}_1^\pm}<104$~GeV.}    
\label{fig:3}  
\end{figure*}  

\begin{figure*}[h!]  
\subfigure[]{  
  \includegraphics[width=0.99\columnwidth]{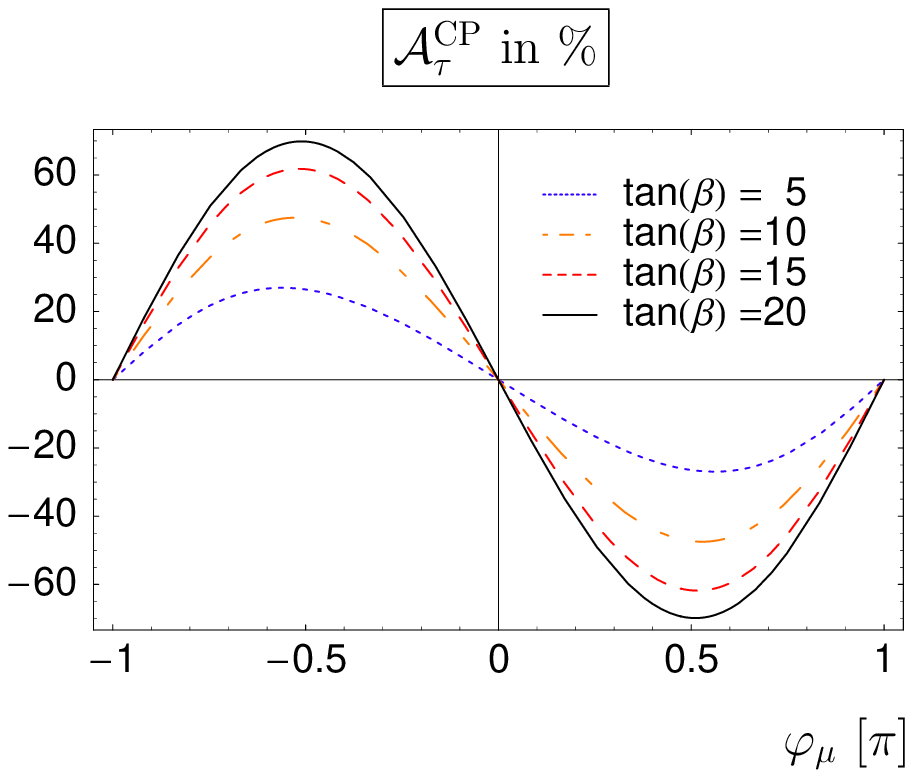}}  
  \hfill  
\subfigure[]{  
  \includegraphics[width=0.99\columnwidth] 
                                       {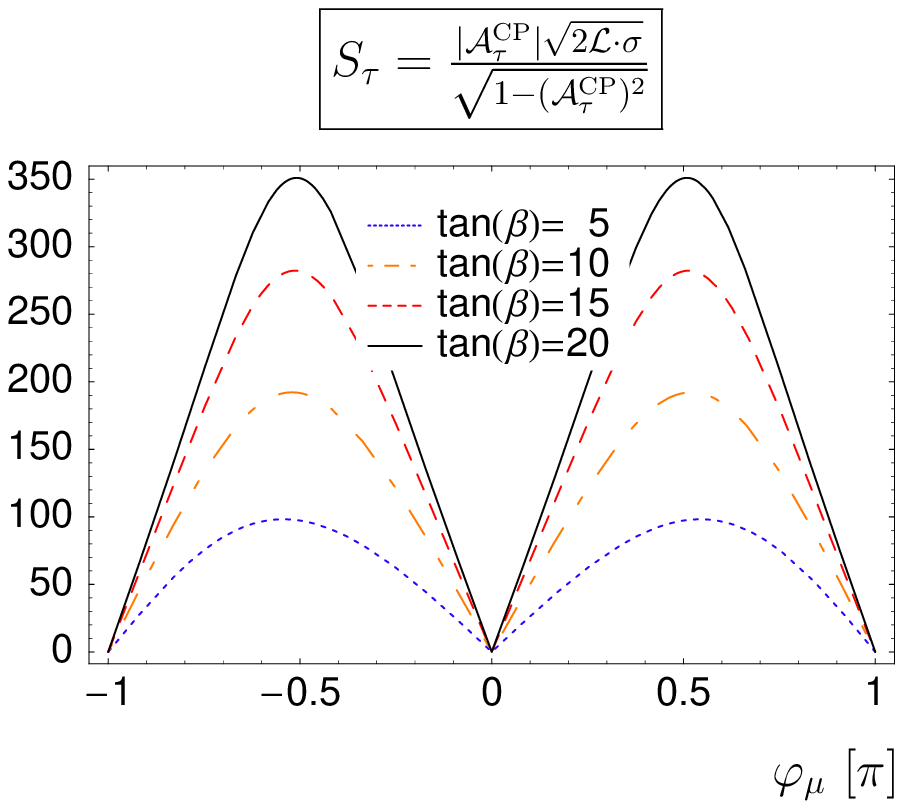}} 
\caption{ 
         Phase dependence of (a) the CP asymmetry of the normal tau polarisation 
         and (b) its  significance for $e^+ e^- \to   
         {\tilde\chi}_1^+ {\tilde\chi}_1^-$; ${\tilde\chi}_1^\pm \to \tau^\pm  
         {\tilde\nu}_\tau^{(\ast)}$, for various values of $\tan\beta$ with  
         $(\mathcal{P}_{e^-}|\mathcal{P}_{e^+})=(-0.8|0.6)$ at 
         $\sqrt{s}=500$~GeV, and   
         $\mathcal{L}=500~{\rm fb}^{-1}$. 
         The other SUSY parameters are defined in Table~\ref{tab:1}. 
          }  
\label{fig:4}   
\end{figure*}

\begin{figure*}[h!]  
\subfigure[]{  
  \includegraphics[width=0.99\columnwidth]{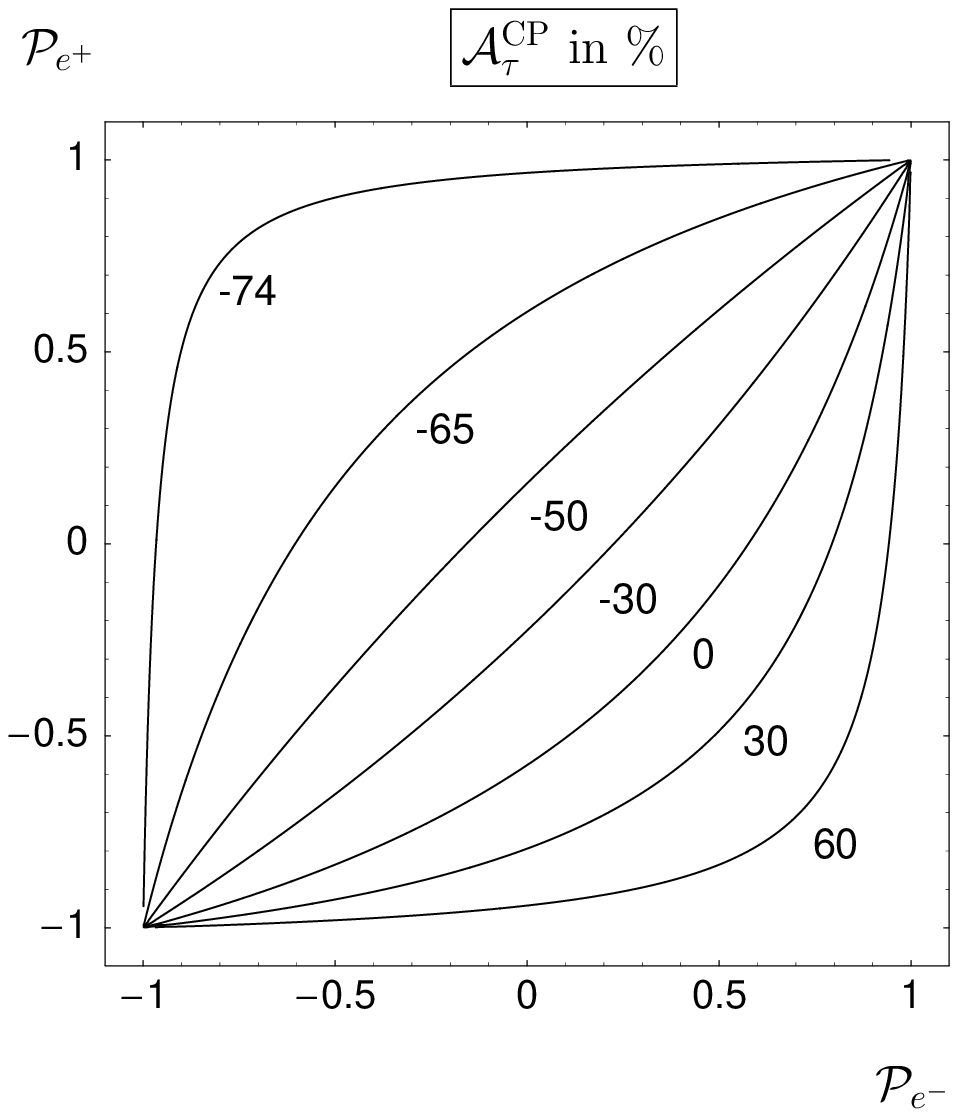}}  
  \hfill  
\subfigure[]{  
  \includegraphics[width=0.99\columnwidth]{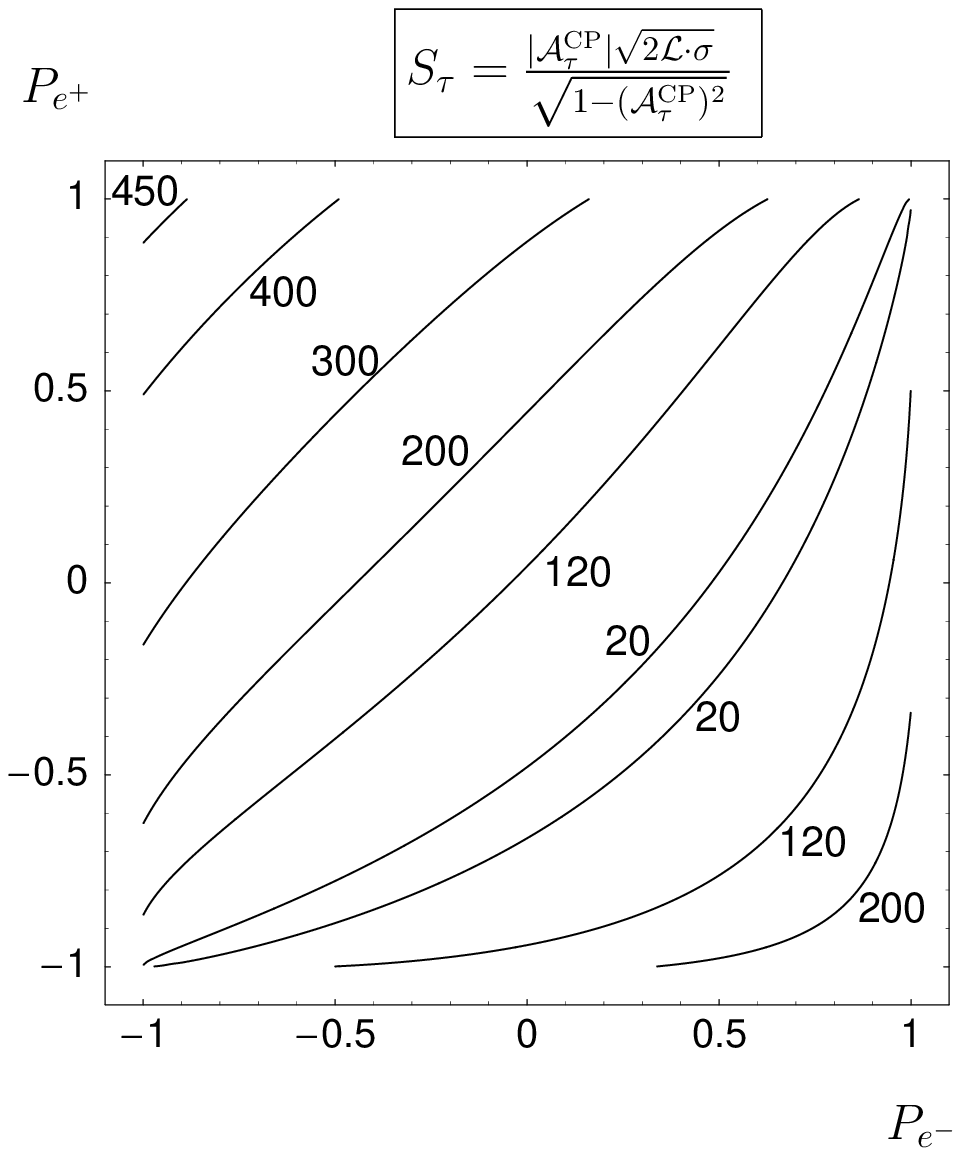}}  
\caption{
         Contour lines in the $\mathcal{P}_{e^+}$--$\mathcal{P}_{e^-}$ plane of
         (a) the CP asymmetry of the normal tau polarisation and  
         (b) its significance for $e^+ e^- \to   
         {\tilde\chi}_1^+ {\tilde\chi}_1^-$; ${\tilde\chi}_1^\pm \to \tau^\pm  
         {\tilde\nu}_\tau^{(\ast)}$,  with  
         $\sqrt{s}=500$~GeV and $\mathcal{L}=500~{\rm fb}^{-1}$, for the  
         scenario in Table~\ref{tab:1}. 
          }    
\label{fig:5}  
\end{figure*}  
\clearpage

is purely real for equal chargino pair
production, ${\tilde\chi}_i^+{\tilde\chi}_i^-$. Then the $Z$-chargino
couplings are real, see Eqs.~\eqref{OprimeL}, \eqref{OprimeR}, and the
t-channel sneutrino ampitude depends only on the modulus of the sneutrino
couplings, $|V_{i1}|^2$, see Eq.~\eqref{eq2}.
Thus, at treelevel, a CP asymmetry in general can only receive CP-odd
contributions from the chargino decay.  We centre our numerical discussion
around a reference scenario, see Table~\ref{tab:1}, which is in some sense
optimised to give large significances. We choose beam polarisations of
$(\mathcal{P}_{e^-}|\mathcal{P}_{e^+})= (-0.8|0.6)$, which enhance the
$\gamma$ exchange and $\gamma Z$ interference contributions in the
production, with respect to the destructive contributions from
$\gamma\tilde\nu_e$ and $Z\tilde\nu_e$ interference. This
favours higher production cross sections and asymmetries compared to
the reversed polarisations $(\mathcal{P}_{e^-}|\mathcal{P}_{e^+})=
(0.8|-0.6)$, since $\gamma Z$ interference then becomes
destructive, too.

\subsubsection{ $M_2$ -- $|\mu|$ dependence}  
\label{subsub:m2mudependence}
In Fig.~\ref{fig:3}(a), we show contour lines of the chargino pair
production cross section $\sigma_P(e^+e^-\to{\tilde\chi}_1^+{\tilde
  \chi}_1^-)$ in the $M_2$--$|\mu|$ plane for the scenario given in
Table~\ref{tab:1}.  For $\mu\lesssim 200$~GeV, the cross section
$\sigma_P$ mainly receives $\gamma$ exchange, and $\gamma Z$
interference contributions, which add to the contributions
from pure $Z$, and $\tilde\nu_e$ exchange, to give more than $\sigma_P
=1000$~fb.  These contributions are about twice as large as the
destructive contributions from $\gamma\tilde\nu_e$ and $Z\tilde\nu_e$
interference.  For $|\mu|, M_2 \gtrsim 200~$GeV, the cross section is
reduced by the growing contributions from $Z\tilde\nu_e$ and
$\gamma\tilde\nu_e$ interference.  For $|\mu|\gtrsim 200~$GeV
and $M_2\lesssim 200~$GeV, the $Z\tilde\nu_e$ and $\gamma\tilde\nu_e$
interference contributions again become weaker, so that the
production cross section is dominated by pure $\tilde\nu_e$ exchange.

\medskip

In Fig.~\ref{fig:3}(b), we show contours for the chargino
branching ratio into the tau, ${\rm BR}({\tilde\chi}_1^+\to\tau^+
\tilde\nu_\tau)$, as a function of $|\mu|$ and $M_2$.  We
indicate the thresholds of the rivalling decay channels by coloured
lines.  The branching ratios ${\rm BR}({\tilde\chi}_1^+\to\ell^+
\tilde\nu_\ell)$ into the light leptons $\ell =e, \mu$ are typically
of the same order as that for the decay into the tau.  Branching
ratios into left sleptons are of the order of ${\rm BR}({\tilde\chi}
_1^+\to{\tilde \ell}_L^+\nu_\ell)<3\%$, $\ell=e,\mu$.  The competitive
chargino decays into staus can reach up to ${\rm BR}({\tilde\chi}_1^+
\to{\tilde\tau}_1^+\nu_\tau)=54\%$ and ${\rm BR}({\tilde\chi}_1^+\to
{\tilde\tau}_2^+\nu_\tau)=15\%$.  Above the cyan-coloured contour in
Fig.~\ref{fig:3}(b), the decay into the $W$ boson opens, with ${\rm
  BR}({\tilde\chi}_1^+\to W^+{\tilde\chi}_1^0) < 5\%$.  The other
chargino decays ${\tilde\chi}_1^+\to W^+{\tilde\chi}^0_{n}$,
$n=2,3,4$, are kinematically excluded, since already
$m_{\tilde\chi^0_2} \approx m_{\tilde\chi^\pm_1}$.

\medskip

In Fig.~\ref{fig:3}(c), we show the CP asymmetry
$\mathcal{A}_\tau^{\rm CP}$, Eq.~(\ref{P2}), within its kinematically
allowed range in the $M_2$--$|\mu|$ plane for chargino production,
$\sqrt{s}\ge 2 m_{\tilde\chi_1^\pm}$, and decay,
$m_{\tilde\chi_1^+}\ge m_{{\tilde\nu}_\tau}$.  For increasing $M_2$,
the asymmetry receives increasing contributions from pure
$\tilde\nu_e$ exchange, whereas the $\gamma\tilde\nu_e$ and
$Z\tilde\nu_e$ interference terms, which enter with opposite sign, get
reduced. Although the cross section $\sigma=\sigma_P(e^+ e^- \to
{\tilde\chi}_1^+ {\tilde\chi}_1^-)\times {\rm BR}({\tilde\chi}_1^+ \to
\tau^+ {\tilde\nu}_\tau)$, which enters the denominator of
Eq.~\eqref{P2}, also increases with increasing $M_2$, the asymmetry
attains its maximum of more than $\mathcal{A}_\tau^{\rm CP}=-70\%$ for
large $M_2\gsim350$~GeV.  The reason is that the coupling factor
$\eta_1$, Eq.~\eqref{etafactor}, to which the asymmetry is
proportional, is here also maximal for large $M_2$.  As discussed in
Subsect.~\ref{sub:analytics}, $\eta_1$ and thus the asymmetry is
largest for $|V_{11}| \sim |Y_\tau U_{12}|$. For $\tan\beta=25$, we
have $Y_\tau \sim 1/3$, resulting in a maximum of $\eta_1$ and the
asymmetry for $|\mu|\sim M_2/2$ see Fig.~\ref{fig:2}(a), which is in
good agreement with the location of the maximum of the asymmetry in
Fig.~\ref{fig:3}(c).

\medskip

In Fig.~\ref{fig:3}(d), we show the corresponding theoretical
significance $S_\tau$, which is defined in Eq.~\eqref{significance},
Appendix~\ref{sec:significance}, for 
$\mathcal{L}=500~$fb$^{-1}$.

\subsubsection{$\varphi_\mu$  and  $\tan\beta$ dependence}  
\label{subsub:phitanbetadependence}
 
In Figs.~\ref{fig:4}(a) and (b), we show the $\varphi_\mu$ dependence
of $\mathcal{A}_\tau^{\rm CP}$ and $S_\tau$, respectively, for
different values of $\tan\beta$.  First, the asymmetry is increasing
for increasing $\tan\beta$, since $\mathcal{A}_\tau^{\rm CP}\sim
Y_\tau$, as discussed in Subsection~\ref{sub:analytics}.  Second,
concerning the phase dependence of the asymmetry, we find for large
$\tan\beta$ that $\mathcal{A}_\tau^{\rm CP}\sim \sin(\varphi_\mu)$, as
also discussed in Subsection~\ref{sub:analytics}.  We observe from
Fig.~\ref{fig:4}(a) the almost perfect sinusoidal behaviour of
$\mathcal{A}_\tau^{\rm CP}$ for large $\tan\beta=20$. For smaller
values of $\tan\beta$, the sine-shape of the asymmetry gets less
pronounced such that its maxima are not necessarily obtained
for maximal CP phases $\varphi_\mu=\pm\pi/2$, but are slightly shifted
away.  It is remarkable that in our scenario, see Table~\ref{tab:1},
the asymmetry can still be $\mathcal{A}_\tau^{\rm CP}=\pm22\%$ even
for $\varphi_\mu=\pm0.1\pi$.  Small phases $\varphi_\mu$ are suggested
by the experimental upper bounds on the EDMs, and the asymmetry will
be a sensitive probe even for small CP phases.

\subsubsection{Beam polarisation dependence}  
             \label{subsub:beampoldependence}

In Fig.~\ref{fig:5}(a), we show the beam polarisation dependence of the   
asymmetry $\mathcal{A}_\tau^{\rm CP}$ for our benchmark scenario, see 
Table~\ref{tab:1}.   
For unpolarised beams the asymmetry is $\mathcal{A}_\tau^{\rm CP}=-43\%$,   
and varies between $\mathcal{A}_\tau^{\rm CP}=-74\%$ for  
$(\mathcal{P}_{e^-}|\mathcal{P}_{e^+})=(-0.8|0.6)$, and $\mathcal{A}_\tau^{\rm CP} =+60\%$ for  
$(\mathcal{P}_{e^-}|\mathcal{P}_{e^+})=(0.8|-0.6)$.  
The strong dependence of the asymmetry on the beam polarisations is due   
to the enhancement of the chargino production channels with $\tilde\nu_e$ 
exchange  for negative electron beam polarisation, $\mathcal{P}_{e^-}<0$,   
and positive positron beam polarisation, $\mathcal{P}_{e^+}>0$.   
For oppositely polarised beams,  $\mathcal{P}_{e^-}>0$, $\mathcal{P}_{e^+}<0$,  
the  $Z$ exchange contributions are enhanced, and those of $\tilde \nu_e$ are  
suppressed. Since the $Z$ contributions enter with opposite sign, also the 
sign   of $\mathcal{A}_\tau^{\rm CP}$ changes, see Fig.~\ref{fig:5}(a).\\

The corresponding theoretical statistical 
significance $S_\tau$ is shown in Fig.~\ref{fig:5}(b).  The production
cross section $\sigma_P(e^+e^-\to {\tilde\chi}_1^+{\tilde\chi}_1^-)$
varies from $332~$fb for unpolarised beams to $418~$fb for
$(\mathcal{P}_{e^-}|\mathcal{P}_{e^+})=(-0.8|0.6)$, and $121~$fb for
$(\mathcal{P}_{e^-}|\mathcal{P}_{e^+})=(0.8|-0.6)$.  Thus the
largest values of $\sigma_P$ are obtained for polarised beams, where
$\tilde \nu_e$ exchange contributions are enhanced. The significance
then reaches up to $S_\tau=450$.

\medskip

\subsection{Production of an unequal pair of charginos  
$\boldsymbol{e^+ +  e^- \to {\tilde\chi}_i^+ + {\tilde\chi}_j^-}$ and  
decay $\boldsymbol{   
           {\tilde\chi}_i^+ \to \tau^+ + {\tilde\nu}_\tau}$  }
\label{sub:12prod}
  
For ${\tilde\chi}_1^\pm{\tilde\chi}_2^\mp$ production, the CP
asymmetry $\mathcal{A}_\tau^{\rm CP}$ in principle also receives
non-vanishing CP-odd contributions from the production.  However, in
our benchmark scenario with large $\tan\beta=25$, see
Table~\ref{tab:2}, those contributions are smaller than $1\%$. The
dominant contributions will still be from the decay, and we
discuss the asymmetries for the decay of ${\tilde\chi}_1^\pm$ and
${\tilde\chi}_2^\pm$ separately.

\medskip

\begin{table}[b]  
\caption{Scenario for $e^+ e^- \to {\tilde\chi}_{1}^\pm 
        {\tilde\chi}_{2}^\mp$ production and decay ${\tilde\chi}_{1,2}^\pm \to  
        \tau^\pm {\tilde\nu}_\tau^{(\ast)}$. The mass parameters $M_2$, $|\mu|$,  
        $M_{\tilde E}$ and $M_{\tilde L}$ are given in GeV.}  
\label{scen12prod}  
\begin{tabular}{cp{0.03\columnwidth}  
                cp{0.07\columnwidth}  
                cp{0.11\columnwidth}  
                cp{0.08\columnwidth}  
                c}  
\\  
\hline\hline  
$\tan\beta$ && $\varphi_\mu$ && $M_2$ && $|\mu|$  &&   
$M_{\tilde E}=M_{\tilde L}$   
\\  
\hline  
25 && $0.5\pi$ && 250 && 200 && 150   
\\  
\hline\hline  
&&&&&&&&  
\\  
\multicolumn{9}{c}{Calculated mass spectrum.}  
\\  
\hline\hline  
${\tilde\ell}$                 && $m$~[GeV]&&&&$\tilde\chi$   &&$m$~[GeV]\\ 
\hline    
${\tilde e}_R,{\tilde \mu}_R$  && 156      &&&&${\tilde\chi}_1^0$&& 115\\  
${\tilde e}_L,{\tilde \mu}_L$  && 157      &&&&${\tilde\chi}_2^0$&& 177\\  
${\tilde\nu}_e,{\tilde\nu}_\mu$&& 136      &&&&${\tilde\chi}_3^0$&& 210\\  
${\tilde\tau}_1$               && 125      &&&&${\tilde\chi}_4^0$&& 294\\  
${\tilde\tau}_2$               && 183      &&&&${\tilde\chi}_1^\pm$&& 170\\ 
${\tilde\nu}_\tau$             && 136      &&&&${\tilde\chi}_2^\pm$&& 294\\ 
\hline\hline  
\multicolumn{3}{c}{${\rm BR}({\tilde\chi}_1^+\to\tau^+\tilde\nu_\tau)$~[\%]}&&&&    
\multicolumn{3}{c}{${\rm BR}({\tilde\chi}_2^+\to\tau^+\tilde\nu_\tau)$~[\%]}\\   
\hline  
                               &&\enspace 28 &&&&           &&\enspace 14\\ 
\hline\hline  
\multicolumn{9}{c}{$\sigma_P(e^+e^-\to  
                   {\tilde\chi}_1^\pm {\tilde\chi}_2^\mp)$~[fb]}\\   
\hline  
\multicolumn{9}{c}{444}\\ 
\hline\hline  
\end{tabular}  
\label{tab:2}  
\end{table}

\subsubsection{Decay of~ $\boldsymbol{{\tilde\chi}_2^+ \to \tau^+ +  
               {\tilde\nu}_\tau}$}  
\label{subsub:21prod2dec}
 
In Fig.~\ref{fig:6}(a), we show the $M_2$--$|\mu|$ dependence of the
production cross section $\sigma_P(e^+e^-\to{\tilde\chi}_2^+{\tilde
  \chi}_1^-)$ which can attain values of several hundred fb. In
contrast to the production of an equal pair of charginos, $e^+e^-\to
{\tilde\chi}_i^+{\tilde\chi}_i^-$, the cross section receives
destructive contributions from $Z\tilde\nu_e$ interference,
only.  The dominant contribution is from pure $\tilde\nu_e$ exchange.
With increasing $|\mu|$, that contribution decreases faster than the
$Z\tilde\nu_e$ interference term and the production cross section is
reduced.

\medskip

In contrast to the lightest chargino $ \tilde\chi_1^+$, the decay of
the heavy chargino $\tilde\chi_2^+ \to \tau^+\tilde\nu_\tau$ is
kinematically allowed in the entire $M_2$--$|\mu|$ plane, see
Fig.~\ref{fig:6}(b). However, its branching ratio is small, $
{\rm BR}({\tilde\chi}_2^+ \to \tau^+\tilde\nu_\tau) < 14\%$, since all
other decay channels can open, except ${\rm BR}({\tilde\chi}_
2^+ \to{\tilde\chi }_4^0W^+)$, see Eq.~(\ref{chardecyas}). Within the
parameter range of Fig.~\ref{fig:6}(b), we find ${\rm BR}({\tilde\chi}
_2^+\to\ell\tilde \nu_\ell) <28\%$, and ${\rm BR}({\tilde\chi}_2^+\to
{\tilde\ell}_L \nu_\ell) < 15\%$ for $\ell = e,\mu$, as well as ${\rm
  BR}({\tilde\chi }_2^+ \to {\tilde\chi}_2^0W^+) < 41\% $.  Other
decays can reach up to ${\rm BR}({\tilde\chi}_2^+ \to{\tilde\chi}_i^0
W^+) =7\%$, for $i=1,3$, and ${\rm BR}({\tilde\chi}_2^+ \to {\tilde
\chi}_1^+H^0_1) =21\%$.\\

\medskip 
   
Fig.~\ref{fig:6}(c) shows the CP asymmetry $\mathcal{A}_\tau^{\rm
  CP}$.  It reaches more than $70\%$, and is enhanced kinematically by
the rising destructive interference $Z\tilde\nu_e$ in the
production process for $|\mu|\gtrsim 200$~GeV. These lead to lower
contributions to ${\rm P}$ and hence to larger asymmetries,
\textit{cf.}  Eq.~\eqref{P2}.  In addition, the coupling
factor $\eta_2$, Eq.~\eqref{etafactor}, is maximal for $|\mu|\approx
300$~GeV and $M_2\approx 200$~GeV, the condition for maximal
interference of the gaugino-higgsino components. See the
discussion in Subsection~\ref{sub:analytics} and Fig.~\ref{fig:2}(b).\\

\medskip

The corresponding significance, $S_\tau$, Eq.~\eqref{significance},
which is shown in Fig.~\ref{fig:6}(d), is smaller than for the
production of an equal pair of charginos, as the cross section
$\sigma_P(e^+e^-\to {\tilde\chi}_2^+{\tilde\chi}_1^-)\times {\rm
  BR}({\tilde\chi}_2^+ \to \tau^+\tilde\nu_\tau)$ is smaller by a
factor of about $2$.  However for $\mathcal{L}=500~$fb$^{-1}$, it can
still attain values of $S_\tau=150$.\\

\medskip
  
\begin{figure*}[h!]  
\subfigure[]{  
  \includegraphics[angle = 0, width=0.96\columnwidth]  
                  {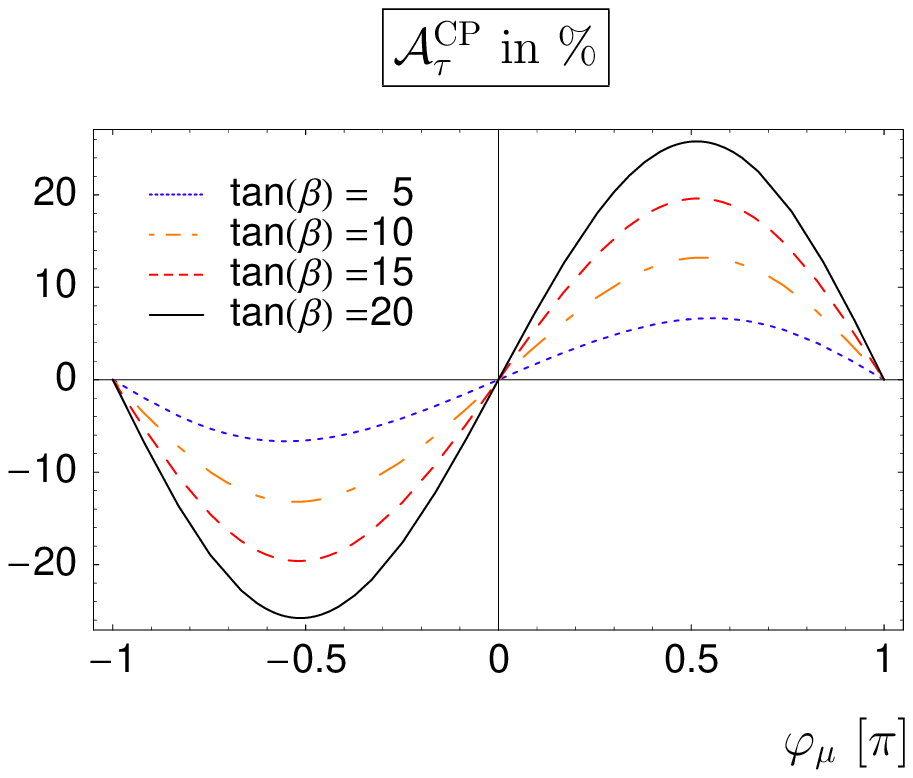}}
\hfill\subfigure[]{  
  \includegraphics[angle = 0, width=0.95\columnwidth]  
                  {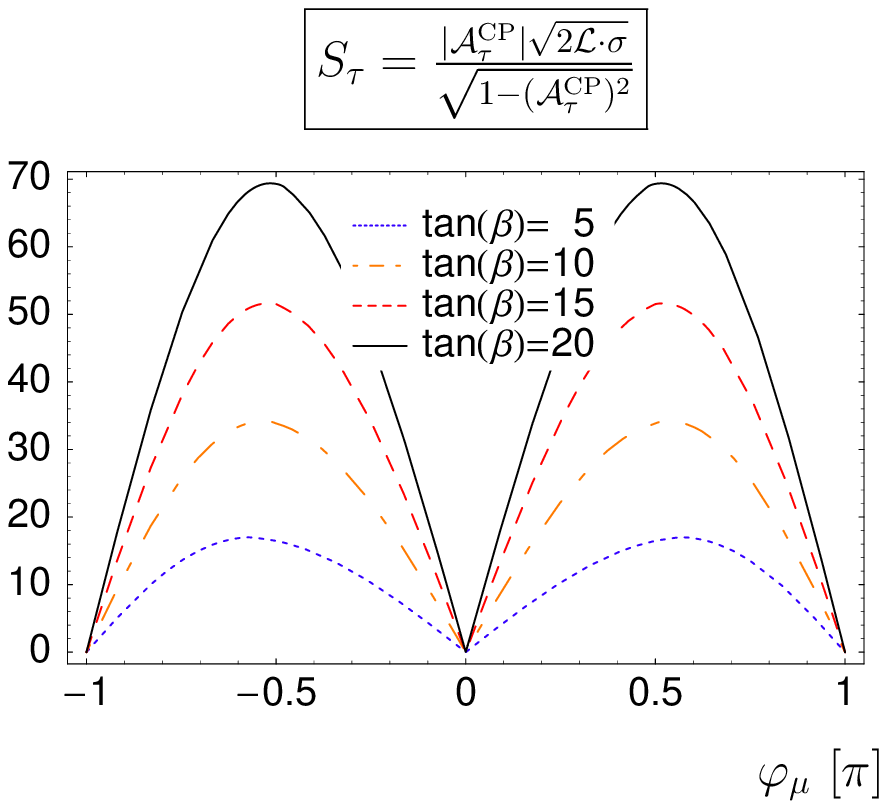}}   
\caption{Phase dependence of (a) the CP asymmetry of the normal tau
         polarisation  and  (b) its significance for $e^+ e^- \to
         {\tilde\chi}_2^\pm {\tilde\chi}_1^\mp$; ${\tilde\chi}_2^\pm \to
         \tau^\pm {\tilde\nu}_\tau^{(\ast)}$, for various values of
         $\tan\beta$ with $(\mathcal{P}_{e^-}|\mathcal{P}_{e^+})=(-0.8|0.6)$
         at $\sqrt{s}=500$~GeV, and $\mathcal{L}=500~{\rm fb}^{-1}$. 
         The other SUSY parameters are defined in Table~\ref{tab:2}. 
         }  
\label{fig:7}    
\end{figure*}

The $\varphi_\mu$ dependence of the asymmetry and its significance   
is shown for various values of $\tan\beta$ in Fig.~\ref{fig:7}. 
Again, we can clearly observe the two striking features, 
$\mathcal{A}_\tau^{\rm CP}\propto Y_\tau$, and 
$\mathcal{A}_\tau^{\rm CP}\propto \sin(\varphi_\mu)$. 
These  are due to the special dependence of the asymmetry 
on the $\tau$--$\tilde\nu_\tau$--chargino couplings, and  
can be qualitatively understood, see the discussion 
 in Subsection~\ref{sub:analytics}.

\medskip

\vfill 
\subsubsection{Decay of $\boldsymbol{{\tilde\chi}_1^+ \to \tau^+ +  
              {\tilde\nu}_\tau}$}   
\label{subsub:12prod1dec}
  
\begin{figure*}[h!]  
\subfigure[]{ 
  \includegraphics[width=0.99\columnwidth]{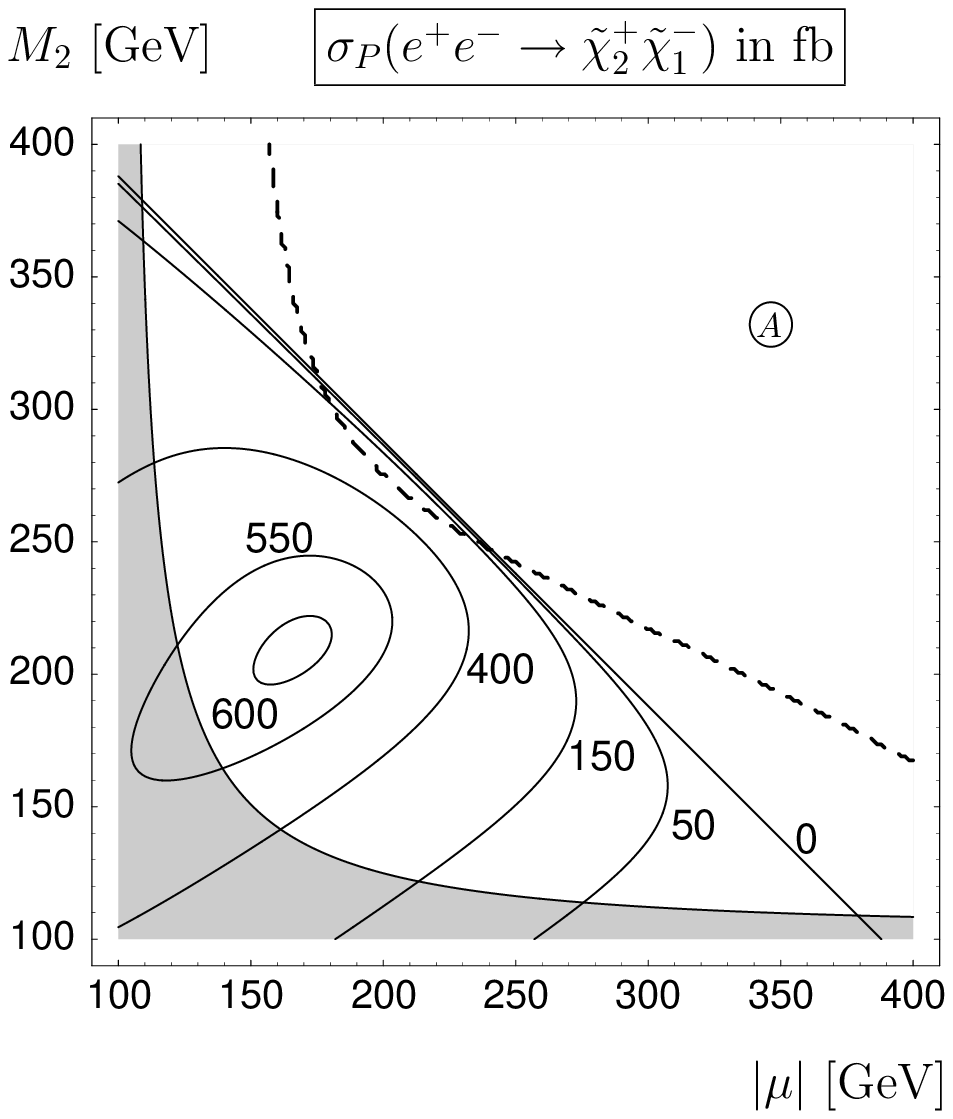}} 
  \hfill  
\subfigure[]{ 
  \includegraphics[width=0.99\columnwidth]{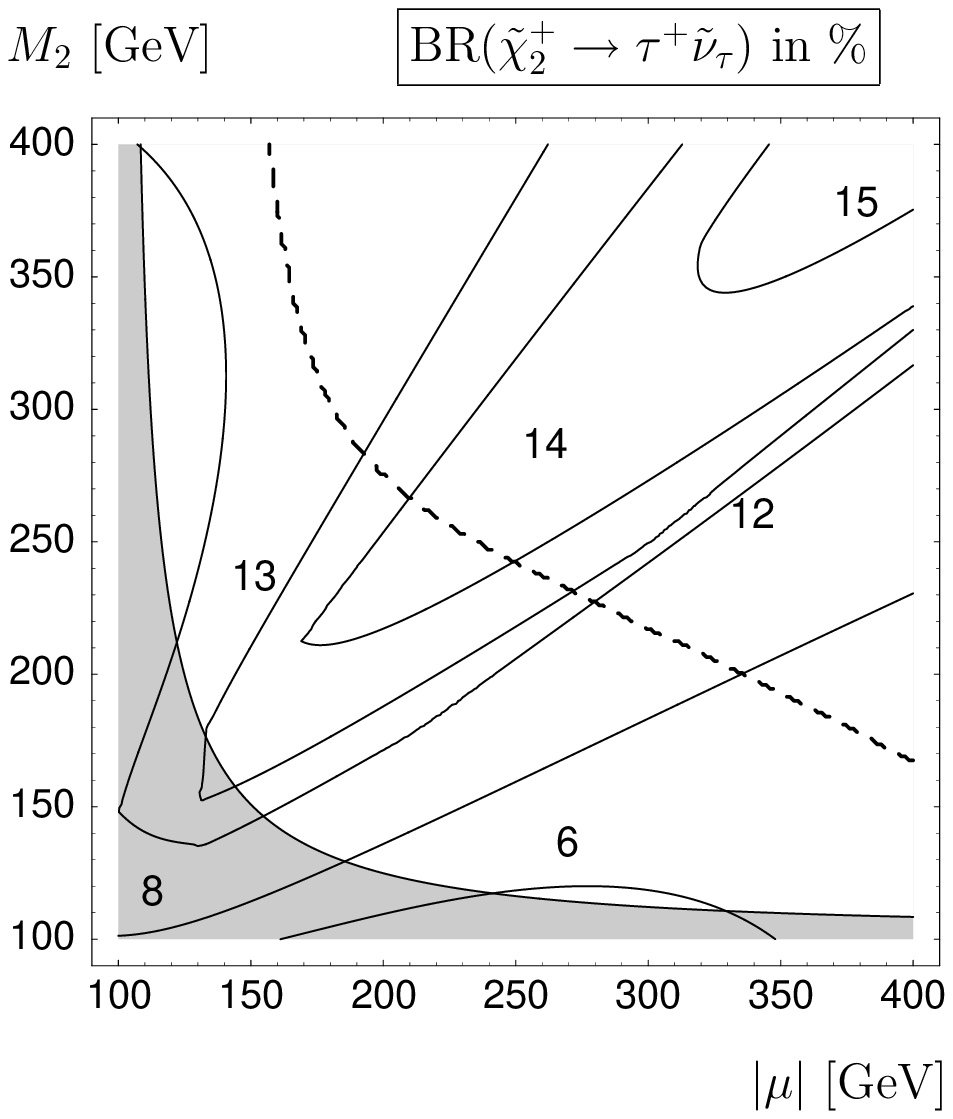}} 
\subfigure[]{   
  \includegraphics[width=0.99\columnwidth]{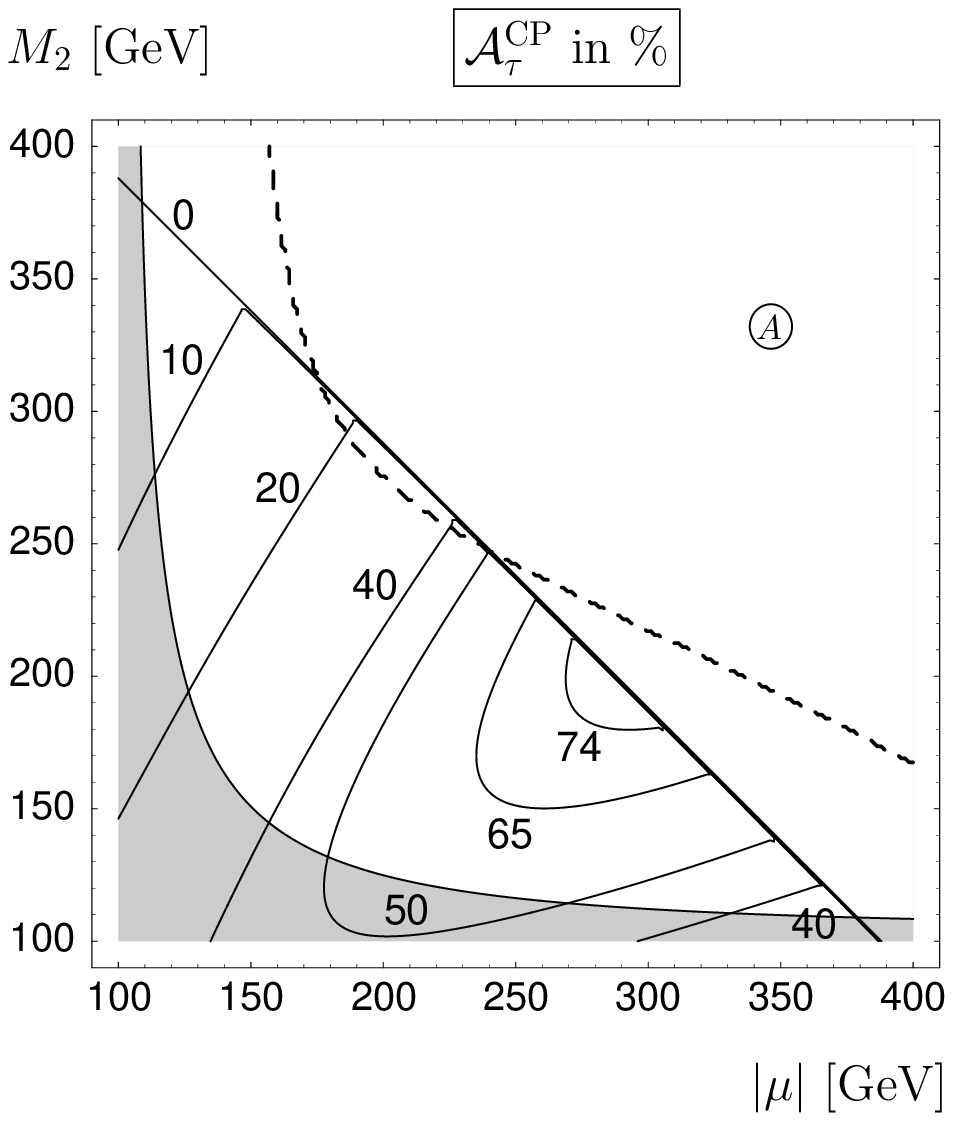}}  
  \hfill  
\subfigure[]{ 
  \includegraphics[width=0.99\columnwidth]{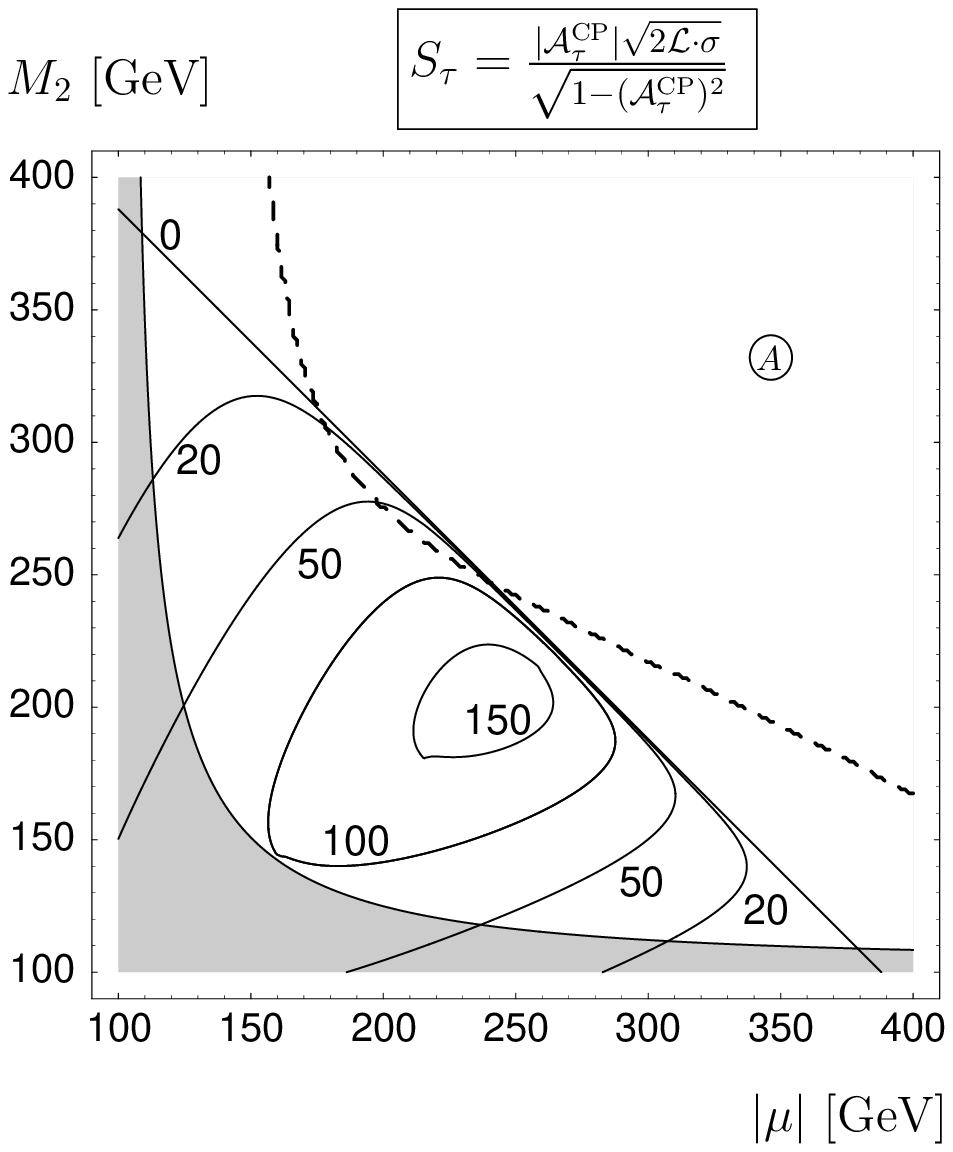}}   
\caption{ 
         Contour lines in the $M_2$-$|\mu|$ plane of (a) the production cross
         section, (b) branching ratio, (c) CP asymmetry  of the normal tau
         polarisation  and (d) its significance for $e^+ e^- \to
         {\tilde\chi}_2^\pm {\tilde\chi}_1^\mp$; ${\tilde\chi}_2^\pm \to
         \tau^\pm  {\tilde\nu}_\tau^{(\ast)}$, with a  centre-of-mass energy
         $\sqrt{s}=500$~GeV, longitudinally polarised beams
         $(\mathcal{P}_{e^-}|\mathcal{P}_{e^+})=(-0.8|0.6)$, and an integrated 
         luminosity $\mathcal{L}=500~{\rm fb}^{-1}$. The other SUSY parameters
         are defined in Table~\ref{tab:2}. The  area
         {\large$\bigcirc$}\hspace*{-0.36cm}{\small $A$}\, above the zero  
        contour line of the cross section is kinematically   
         forbidden by $\sqrt{s}< m_{{\tilde\chi}_2^+}+m_{{\tilde\chi}_1^-}$. 
         Above the dashed line the lightest neutralino is no longer the LSP
         since $m_{{\tilde\tau}_1}<{\tilde\chi}_1^0$. In the grey-shaded area  
         $m_{{\tilde\chi}_1^\pm}<104$~GeV.  
         }   
\label{fig:6}    
\end{figure*}

\begin{figure*}[h!]  
\subfigure[]{   
  \includegraphics[width=0.94\columnwidth]{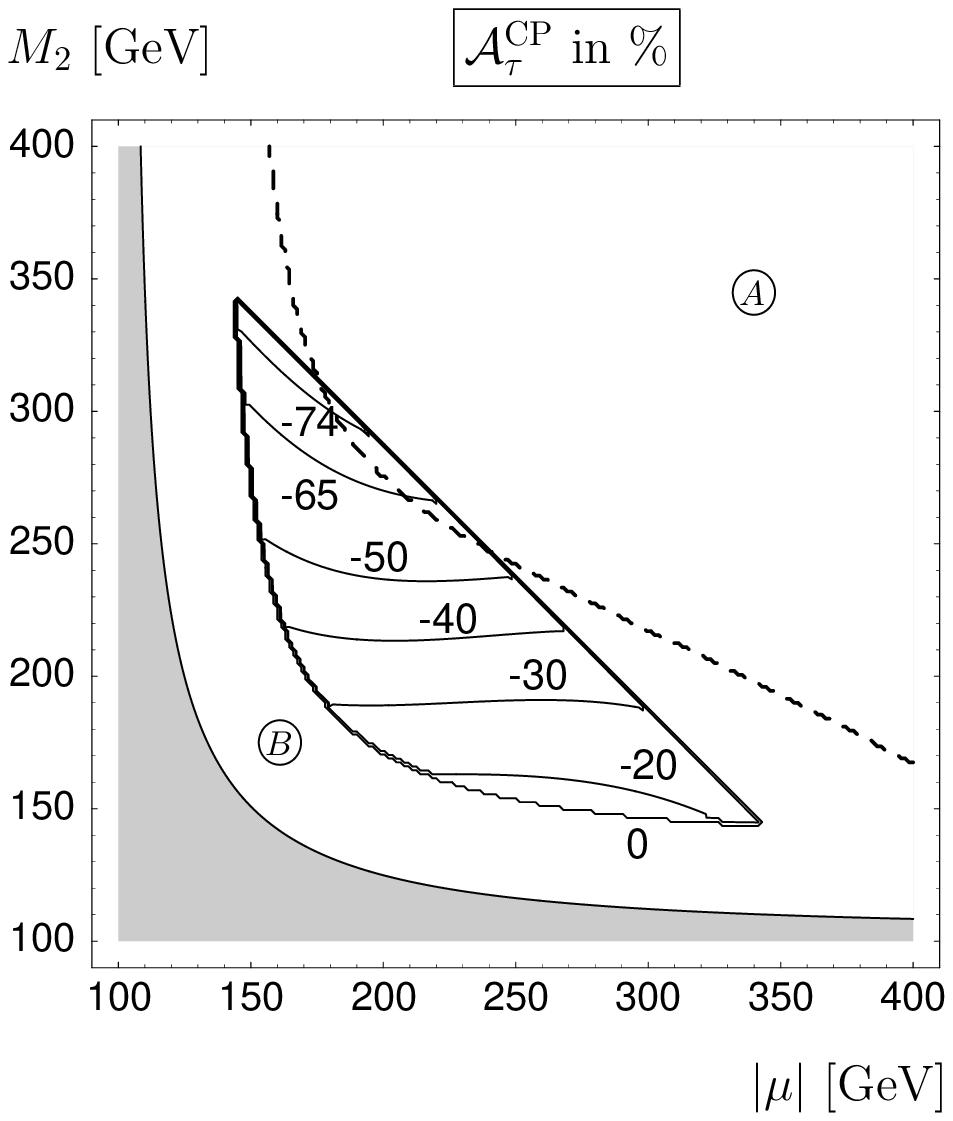}}  
\hfill
\subfigure[]{  
  \includegraphics[width=0.94\columnwidth]{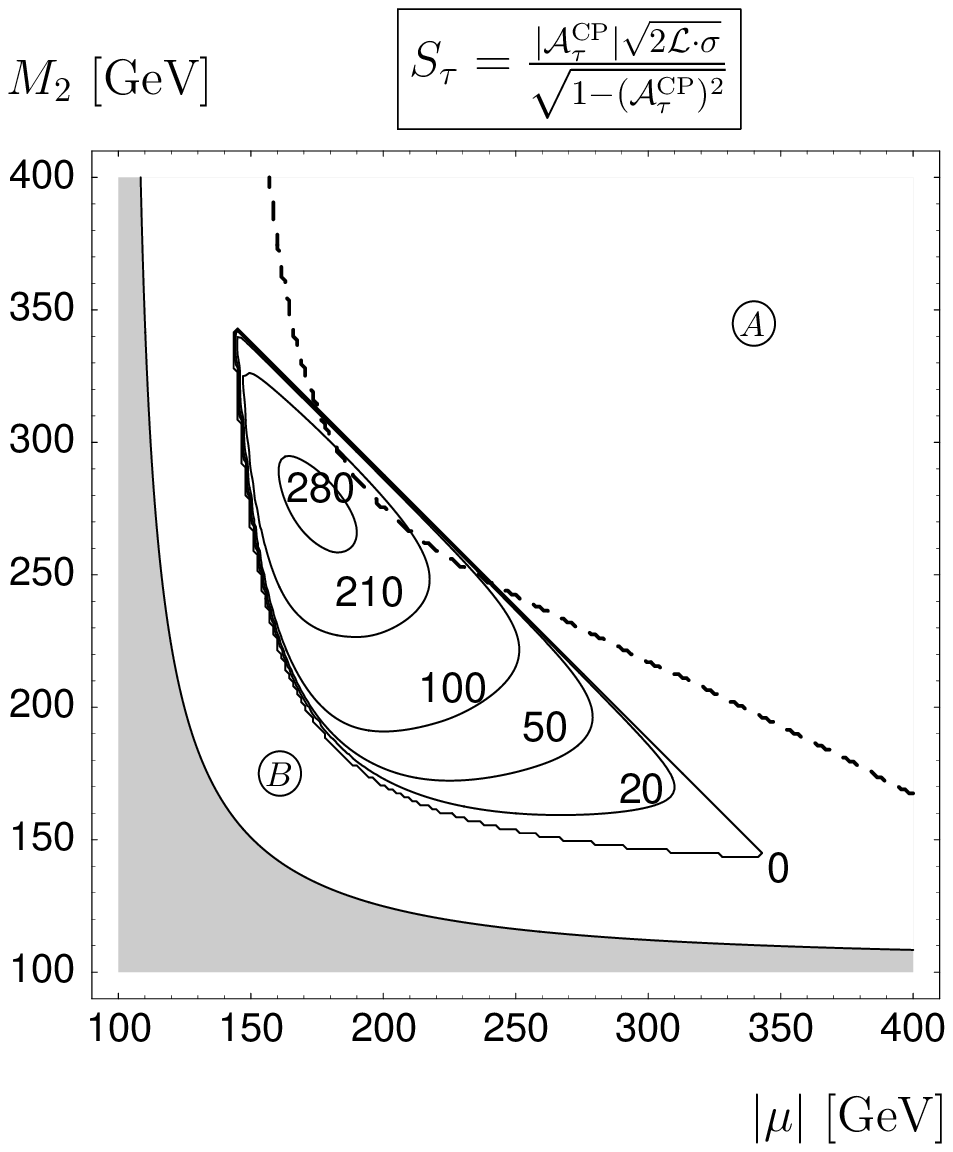}}  
\caption{ 
         Contour lines in the $M_2$-$|\mu|$ plane of  
         (a) the  CP asymmetry of the  normal tau polarisation and (b) its  
         significance for $e^+ e^- \to {\tilde\chi}_1^\pm {\tilde\chi}_2^\pm$; 
         ${\tilde\chi}_1^\pm \to \tau^\pm {\tilde\nu}_\tau^{(\ast)}$,  with a  
         centre-of-mass energy $\sqrt{s}=500$~GeV, longitudinally polarised
         beams $(\mathcal{P}_{e^-}|\mathcal{P}_{e^+})=(-0.8|0.6)$, and  an
         integrated luminosity $\mathcal{L}=500~{\rm fb}^{-1}$. The other SUSY
         parameters are defined in Table~\ref{tab:2}. 
         The area {\large$\bigcirc$}\hspace*{-0.37cm}{\small  
         $A$}\, is kinematically forbidden by $\sqrt{s}< m_{{\tilde\chi}_1^+}+ 
         m_{{\tilde\chi}_2^-}$,  and the area 
         {\large$\bigcirc$}\hspace*{-0.37cm}{\small $B$}\,  
          is kinematically forbidden by  
         $m_{{\tilde\chi}_1^+}< m_{{\tilde\nu}_\tau}$. Above the   
         dashed line the lightest neutralino is no longer the LSP since  
          $m_{{\tilde\tau}_1} <{\tilde\chi}_1^0$. In the   
         grey-shaded area $m_{{\tilde\chi}_1^\pm}<104$~GeV.}  
\label{fig:8}    
\end{figure*}

In Fig.~\ref{fig:8}(a), we show the $M_2$--$|\mu|$ dependence of the
CP asymmetry.  Large values, up to $\mathcal{A}_\tau^{\rm CP}=-74\%$,
are obtained towards $M_2 \sim 2|\mu|$, where also $\eta_1$,
Eq.~\eqref{etafactor}, is maximal; compare with the asymmetry in
Fig.~\ref{fig:3}(c).  The corresponding branching ratio ${\rm
  BR}({\tilde\chi}_1^+\to \tau^+ {\tilde\nu}_\tau)$ does not exceed
$30\%$. The decay channels into the light leptons ${\rm
  BR}({\tilde\chi}_1^+\to \ell^+ {\tilde\nu}_\ell)$, $\ell = e,\mu$
and into the lightest stau ${\rm BR}({\tilde\chi}_1^+\to
{\tilde\tau}_1^+ \nu_\tau)$ are the most competitive ones with up to
$20\%$ and $30\%$, respectively.  Towards the production threshold,
${\rm BR}({\tilde\chi}_1^+\to {\tilde \ell}_L \nu_\ell)$ is of the
order of $5\%$. Together with the production cross section
$\sigma_P(e^+e^-\to {\tilde\chi}_1^+{\tilde\chi}_2^-)$, as shown in
Fig.~\ref{fig:6}(a), the product of production and decay
branching ratio, $\sigma=\sigma_P\times {\rm
  BR}({\tilde\chi}_1^+ \to \tau^+\tilde\nu_\tau)$, can be as large as
$140$~fb.  The statistical significance, shown in Fig.~\ref{fig:8}(b),
reaches $S_\tau=200$, for $\mathcal{L}=500~{\rm fb}^{-1}$.

\clearpage

\subsection{Inverted hierarchy scenario 
}\label{sub:super}
 
The phase $\varphi_\mu$ of the higgsino mass parameter $\mu$
contributes already at the one-loop level to the electric dipole
moments (EDMs) of the electron~\cite{Commins:1994gv}, the
neutron~\cite{Baker:2006ts}, and the mercury
atom~\cite{Griffith:2009zz,Falk:1999tm}. The dominant contribution to
the electron EDM from chargino exchange, for example, is proportional
to $\sin\varphi_\mu$~\cite{Commins:1994gv,cancellations1}. The phase
$\varphi_\mu$ is thus strongly constrained by the experimental upper
limits on the EDMs with $|\varphi_\mu|\lesssim 0.1\pi$ in
general~\cite{cancellations2}.  However, the bounds from the EDMs are
highly model dependent. For instance, cancellations between different
SUSY contributions to the EDMs can resolve the restrictions on the
phases~\cite{cancellations1}, although a proper fine tuning of SUSY
parameters is required. On the other hand, the bounds on the phases
might disappear if lepton flavour violation is
included~\cite{cancellations3}.

\medskip

Another way to fulfil the EDM bounds is to assume sufficiently heavy
sleptons and/or squarks. Since sparticle masses of the order of
$10$~TeV are required~\cite{cancellations1}, such solutions are
unnatural. Models with heavy sparticles have been discussed in the
literature, like \emph{split-SUSY}~\cite{ArkaniHamed:2004yi} or
\emph{focus-point} scenarios~\cite{Feng:1999zg}.  If only the first
and second generation squarks are  heavy, naturalness can be
reconciled while experimental constraints can still be fulfilled, see
e.g. Ref.~\cite{Bagger:1999ty} for such an
\emph{inverted hierarchy} approach.

\medskip

Heavy sfermions of the first and second generations are particularly
interesting for our process of chargino production
$e^+e^-\to\tilde\chi_i^\pm\tilde\chi_j^\mp$ and decay into the tau,
$\tilde\chi_i^\pm\to\tau^\pm \tilde\nu_\tau^{(\ast)}$.  First, the
negative contributions from sneutrino $\tilde\nu_e$ interference to
the production cross sections are considerably reduced. Second, the
chargino branching ratio into the tau is enhanced, since the chargino
decay channels $\tilde\ell_L \nu_\ell$ and $\ell \tilde\nu_\ell$ are
closed due to the heavy sleptons of the first and second generations,
$\ell = e, \mu$.  In order to compare with our previous results, we
take the parameters as in Table~\ref{tab:1}, but now choose heavy soft
breaking masses for the selectrons and smuons $M_{\tilde L} =
M_{\tilde E} = 15$~TeV. See the new reference scenario and the
resulting mass spectrum in Table~\ref{tab:3}.

\begin{table}[h!] 
\caption{Scenario for chargino pair production 
  $e^+e^-\to{\tilde\chi}_1^+{\tilde\chi}_1^-$ and decay 
  ${\tilde\chi}_1^\pm\to\tau^\pm{\tilde\nu}_\tau^{(\ast)}$ with heavy first and second 
  slepton generations. The mass parameters $M_2$, $|\mu|$, $M_{\tilde E}$, 
  $M_{{\tilde E}_\tau}$, $M_{\tilde L}$ and $M_{{\tilde L}_\tau}$ are given 
in 
  GeV.} 
\label{tab:3} 
\begin{tabular}{cp{0.003\columnwidth} 
                cp{0.07\columnwidth} 
                cp{0.02\columnwidth} 
                cp{0.03\columnwidth} 
                cp{0.00\columnwidth} 
                c} 
\\ 
\hline\hline 
$\tan\beta$ && $\varphi_\mu$ &\multicolumn{2}{c}{$M_2$} && $|\mu|$ && 
$M_{\tilde E} = M_{\tilde L}$ && 
$M_{\tilde E_\tau}=M_{\tilde L_\tau}$ 
\\ 
\hline 
$25$ && $0.5\pi$ &\multicolumn{2}{c}{$380$} && $240$  && 
$ 15\!\times\!10^3$  && 
$200$ 
\\ 
\hline\hline 
&&&&&&&&& 
\\ 
\multicolumn{11}{c}{Calculated mass spectrum.} 
\\ 
\hline\hline 
${\tilde\ell}$              && $m$~[GeV]&&&&&&$\tilde\chi$  &&$m$~[GeV]\\ 
\hline 
${\tilde e}_R,{\tilde \mu}_R$&& $15\!\times\!10^3$&&&&&& 
${\tilde\chi}_1^0$&& $175$\\ 
${\tilde e}_L,{\tilde \mu}_L$&& $15\!\times\!10^3$&&&&&& 
${\tilde\chi}_2^0$&& $238$\\ 
${\tilde\nu}_e,{\tilde\nu}_\mu$&&$15\!\times\!10^3$&&&&&& 
${\tilde\chi}_3^0$&& $247$\\ 
${\tilde\tau}_1$               && $177$             &&&&&& 
${\tilde\chi}_4^0$&& $405$\\ 
${\tilde\tau}_2$               && $230$             &&&&&& 
${\tilde\chi}_1^\pm$&& $225$\\ 
${\tilde\nu}_\tau$             && $189$              &&&&&& 
${\tilde\chi}_2^\pm$&& $405$\\ 
\hline\hline 
\multicolumn{4}{c}{${\rm BR}({\tilde\chi}_1^+\to\tau^+\tilde\nu_\tau)$~[\%]\qquad}&&& 
\multicolumn{5}{r}{\enspace\enspace\enspace\enspace$\sigma_P(e^+e^-\to 
                   {\tilde\chi}_1^-{\tilde\chi}_1^+)$~[fb]}\\ 
\hline 
                              &&\enspace 49&&&&&&&&805\\ 
\hline\hline 
\end{tabular} 
\end{table} 
 
\begin{figure*}[h!]
\subfigure[]{  
  \includegraphics[width=0.99\columnwidth]{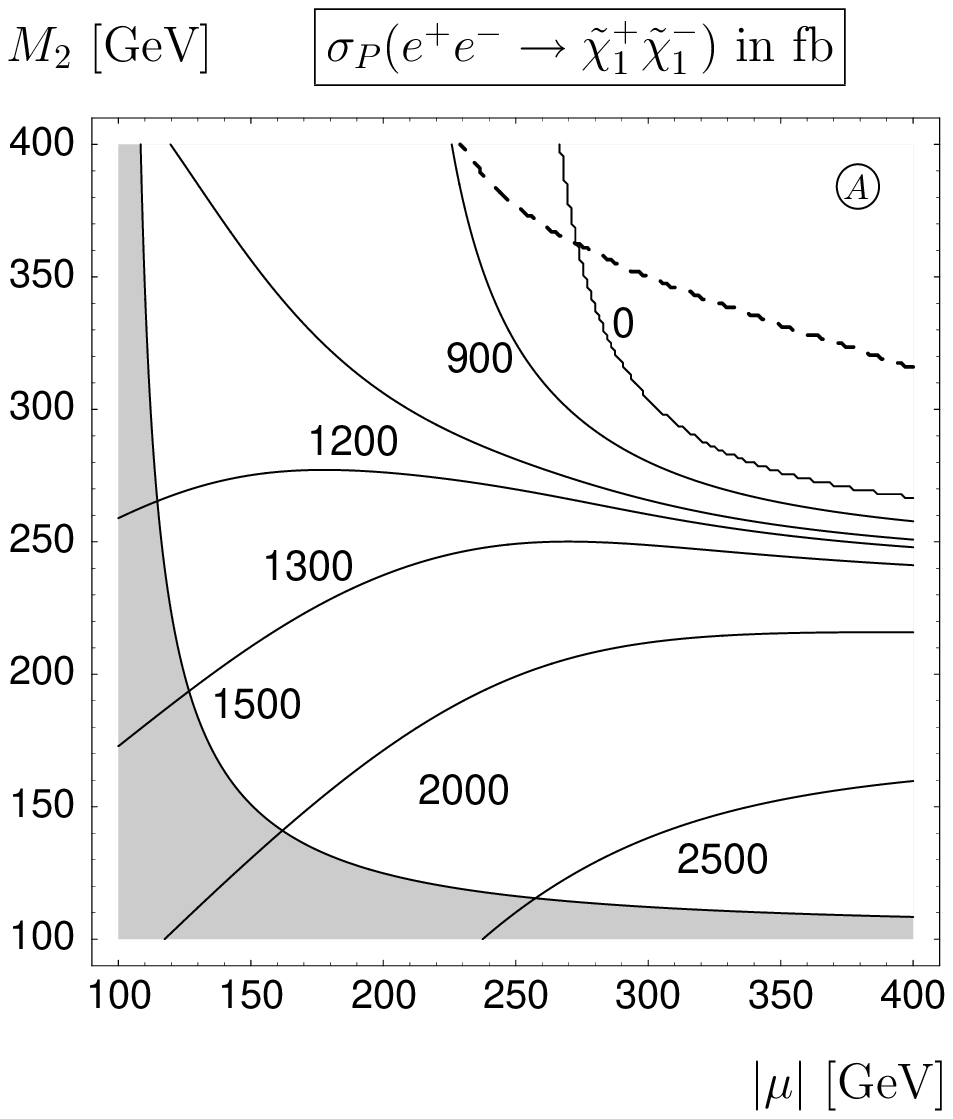} }
  \hfill 
\subfigure[]{
  \includegraphics[width=0.99\columnwidth]{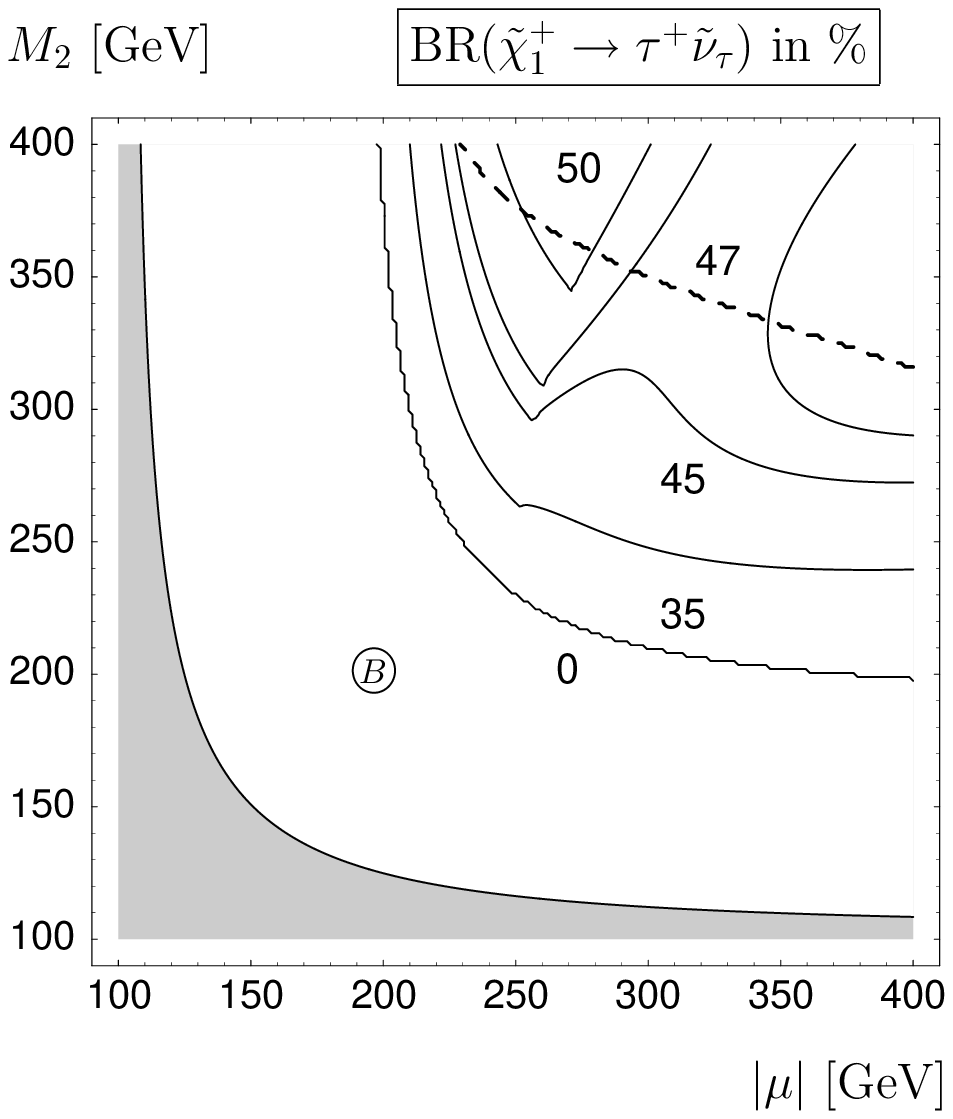} }
\subfigure[]{
  \includegraphics[width=0.99\columnwidth]{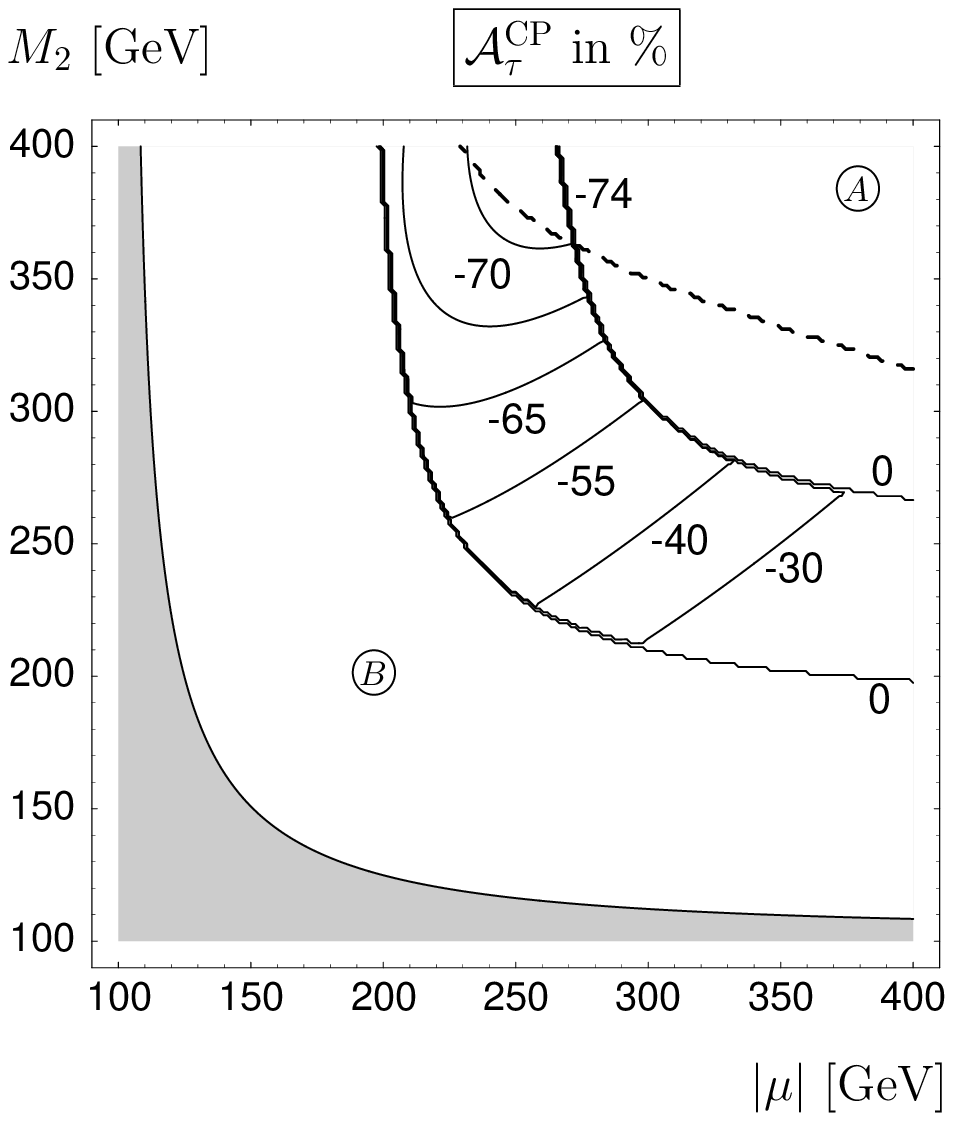} }
  \hfill 
\subfigure[]{
  \includegraphics[width=0.99\columnwidth]{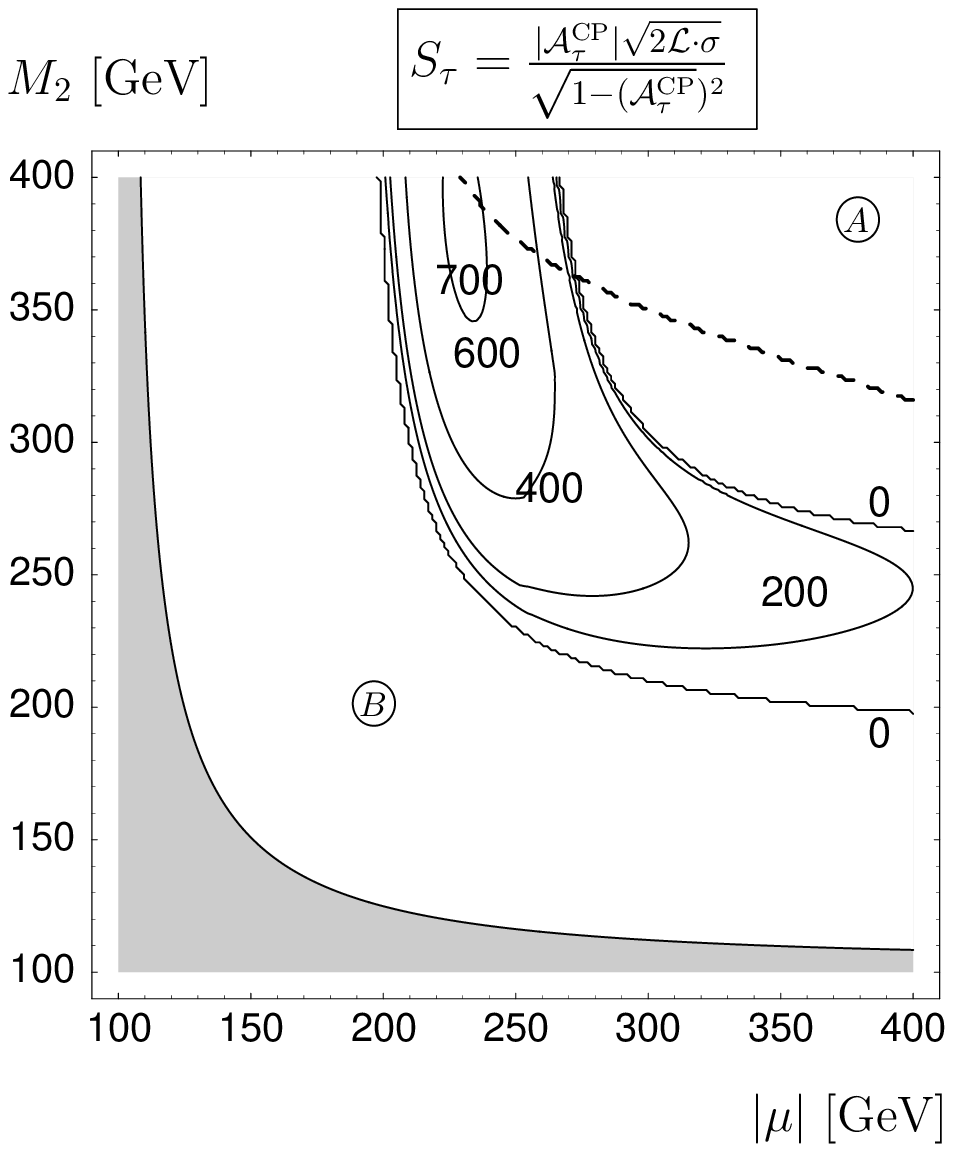} }
\caption{ 
         Contour lines  in the $M_2$--$|\mu|$ plane of (a)  
         the production cross section, (b) branching ratio, 
         (c) CP asymmetry of the normal tau polarisation, and  
         (d) its significance, for 
          $e^+ e^- \to {\tilde\chi}_1^+ {\tilde\chi}_1^-$, ${\tilde\chi}_1^\pm
         \to \tau^\pm {\tilde\nu}_\tau^{(\ast)}$, for  a spectrum of heavy 1st
         and 2nd slepton generations as given in Table~\ref{tab:3}, with a   
         centre-of-mass energy $\sqrt{s}=500$~GeV, longitudinally polarised
         beams $(\mathcal{P}_{e^-}|\mathcal{P}_{e^+})=(-0.8|0.6)$, and an
         integrated luminosity $\mathcal{L}=500~{\rm fb}^{-1}$. The area 
         {\large$\bigcirc$}\hspace*{-0.35cm}{\small $A$}\,\, above the zero 
         contour line of the production cross section is 
         kinematically  forbidden by $\sqrt{s}< 2 m_{{\tilde\chi}_1^\pm} $,   
         and the area  {\large$\bigcirc$}\hspace*{-0.35cm}{\small $B$}\,\, 
         below the  zero contour line of the branching ratio  by  
         $m_{{\tilde\chi}_1^+}< m_{{\tilde\nu}_\tau}$. Above the dashed line 
         the lightest neutralino is no longer the LSP since
         $m_{{\tilde\tau}_1}<{\tilde\chi}_1^0$. In the grey-shaded area  
         $m_{{\tilde\chi}_1^\pm}<104$~GeV} 
\label{fig:9} 
\end{figure*}
 
\begin{figure}[h!] 
\subfigure[]{ 
    \includegraphics[width=0.99\columnwidth]{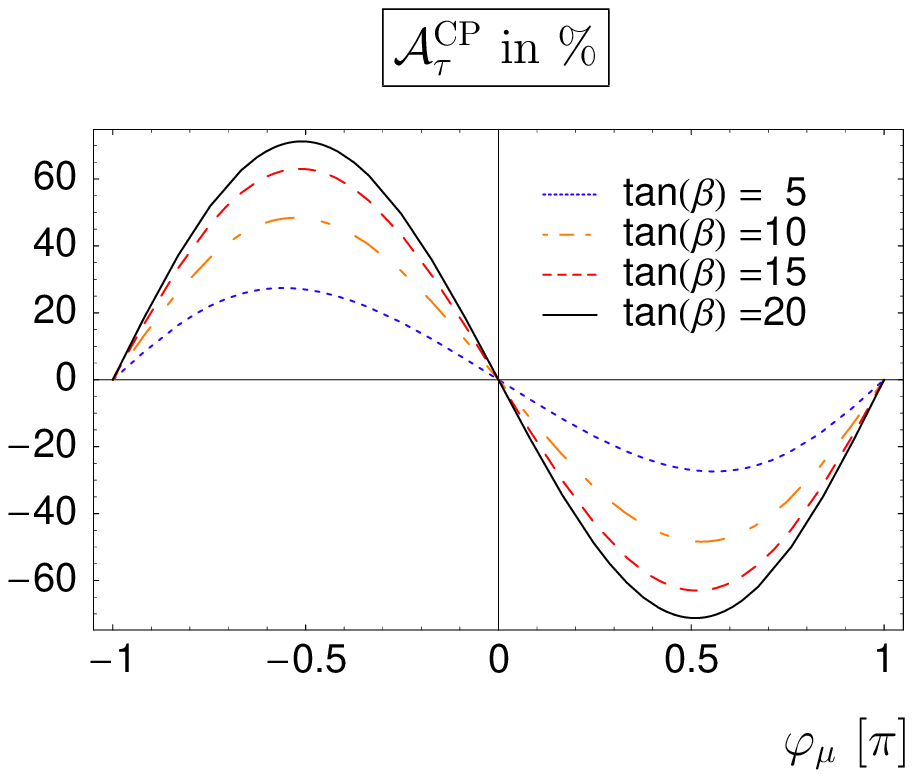}} 
    \phantom{space}\\ 
 
    \phantom{space}
 
\subfigure[]{ 
     \includegraphics[width=0.99\columnwidth]{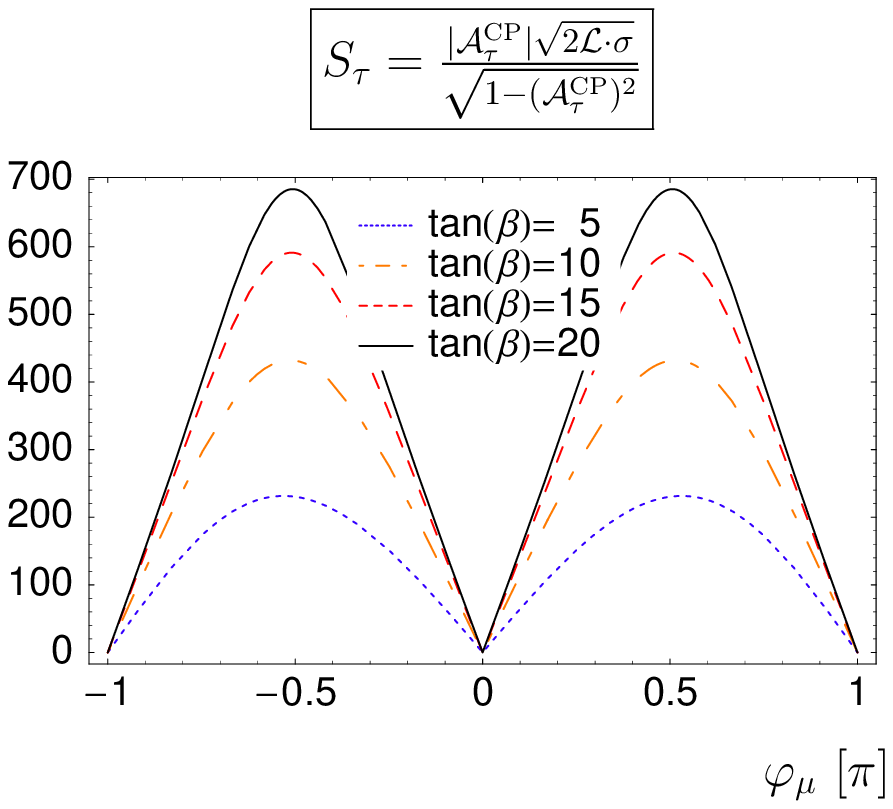}} 
\caption{  
         Phase dependence of (a)  the CP asymmetry of the normal tau
         polarisation,  and (b) its 
         significance  for $e^+ e^- \to {\tilde\chi}_1^+ {\tilde\chi}_1^-$; 
         ${\tilde\chi}_1^\pm \to \tau^\pm {\tilde\nu}_\tau^{(\ast)}$, 
          for  a spectrum of heavy 1st and 2nd slepton generations as given in
         Table~\ref{tab:3}, for various values of  $\tan\beta$, with
         $(\mathcal{P}_{e^-}|\mathcal{P}_{e^+})=(-0.8|0.6)$ at  
         $\sqrt{s}=500$~GeV, and  $\mathcal{L}=500~{\rm fb}^{-1}$. 
         } 
\label{fig:10} 
\end{figure} 
 
\subsubsection{Parameter dependence} 
\label{subsub:superm2mudependence}
 
In Fig.~\ref{fig:9}(a), we show the $M_2$--$|\mu|$ dependence of the
cross section for chargino production
$e^+e^-\to\tilde\chi_1^+\tilde\chi_1^-$ for our new reference
scenario with heavy sneutrinos.  Due to the heavy electron
sneutrino, $m_{{\tilde \nu}_e} = 15$~TeV, the negative interference
contributions from $Z\tilde\nu_e$ and $\gamma\tilde\nu_e$ are
thoroughly suppressed, which enhances the cross section.  In the
scenario with light sneutrinos, in particular for large values of
$|\mu|$, the pure $\tilde\nu_e$ exchange is the largest contribution
to the cross section, see the discussion concerning
Fig.~\ref{fig:3}(a) in Subsection~\ref{subsub:m2mudependence}.
Although the constructive $\tilde\nu_e$ exchange contributions are
also lost for heavy sneutrinos, there is still a net surplus in the
production cross section, compare Fig.~\ref{fig:3}(a) and
Fig.~\ref{fig:9}(a), since they are of the same order as one of the
destructive channels, $Z\tilde\nu_e$ or $\gamma\tilde\nu_e$.

\medskip

The branching ratio ${\rm BR}(\tilde\chi_1^+\to \tau^+
\tilde\nu_\tau)$ is only reduced by the rivaling decay into the
lightest stau, which is at least ${\rm BR}(\tilde\chi_1^+\to\tilde
\tau_1^+\nu_\tau)=50 \%$ in Fig.~\ref{fig:9}(b). The product
of production and decay $\sigma=\sigma_P(e^+e^-\to\tilde\chi_1^+\tilde
\chi_1^-)\times {\rm BR}(\tilde\chi_1^+\to \tau^+ \tilde\nu_\tau)$ for
heavy sneutrinos is thus of the order of several hundred fb, see
Table~\ref{tab:3}.  In contrast to the strong impact of a heavy
sneutrino on the cross section, the CP asymmetry $\mathcal{A}_\tau^
{\rm CP}$ is only slightly enhanced, compare Fig.~\ref{fig:9}(c) with
Fig.~\ref{fig:3}(c). The asymmetry is mainly determined by the
coupling factor $\eta_1$, see Eq.~(\ref{etafactor}), which still
allows for asymmetries of more than $70$\%. Together with the enhanced
cross section, this leads to sizable significances of the order of
several hundred standard deviations over statistical fluctuations,
which we show in Fig.~\ref{fig:9}(d).  Also the phase $\varphi_\mu$
and $\tan\beta$ dependence of the asymmetries is still governed by the
coupling factor $\eta_1$, Eq.~(\ref{etafactor}).  In
Fig.~\ref{fig:10}(a), we observe the same sinus-like dependence of
$\mathcal{A}_\tau^{\rm CP}$, which increases with increasing
$\tan\beta$, cf. Fig.~\ref{fig:4}(a), and see the discussion in
Subsection~\ref{sub:analytics}.

\medskip

To summarise, the CP asymmetries in the decay of a chargino into 
a polarised tau are a powerful tool to probe  $\varphi_\mu$, which 
might be large in particular in scenarios with flavour
violation~\cite{cancellations3}, or heavy sfermions of the first and second
generations~\cite{Bagger:1999ty}.

\section{Summary and conclusions}\label{sec:summary}
 
We have studied CP violation in chargino production with
longitudinally polarised beams, $e^+e^-\to\tilde\chi_i^\pm\tilde\chi_j
^\mp$, and the subsequent two-body decay of one of the charginos into
a tau and a sneutrino, $\tilde\chi_i^\pm \to \tau^\pm\tilde\nu_\tau^
{(\ast)}$.  We have given full analytical expressions for the
amplitude squared, taking into account the complete spin correlations
between production and decay.  We have defined a CP-odd asymmetry
$\mathcal{A}_\tau^{\rm CP}$, which is the normal tau polarisation, and
which is sensitive to the CP phase $\varphi_\mu$ of the
higgsino mass parameter $\mu$.
 
\medskip
 
In a numerical discussion we have considered equal chargino
pair production $e^+e^-\to{\tilde\chi}_1^+{\tilde\chi}_1^-$, and
unequal chargino pair production,
$e^+e^-\to{\tilde\chi}_1^\pm{\tilde\chi}_ 2^\mp$, with the ensuing
decay of either ${\tilde\chi}_1^\pm$ or ${\tilde\chi}_2^\pm$,
respectively. We have studied the dependence of the CP asymmetries on
the MSSM parameters of the chargino sector $M_2,|\mu|,\varphi_\mu,$
and $\tan\beta$.  The asymmetries are considerably enhanced for large
$\tan\beta$, where the tau Yukawa coupling is enhanced.  The size of
the asymmetries also strongly depends on the gaugino-higgsino
composition of the charginos, and can be maximal for equally sized
left and right $\tau$--$\tilde\nu_\tau$--chargino couplings.

\medskip

We have found that $\mathcal{A}_\tau^{\rm CP}$ can attain values of more than 
$70\%$. The asymmetry is already present at tree level and can be sizable even 
for small phases of $\mu$, as suggested by the experimental limits on EDMs.  
Moreover, by choosing different beam polarisations the $Z$, $\gamma$ and 
$\tilde\nu_e$ contributions can be enhanced or suppressed. A proper choice of 
beam polarisations can thus considerably enhance both, the asymmetry, and the 
production cross sections. An analysis of statistical errors 
shows that the asymmetries are well accessible in future $e^+e^-$ collider 
experiments in the $500$~GeV range with high luminosity and longitudinally 
polarised beams. 

\medskip

Since the phase $\varphi_\mu$ of the higgsino mass paramter  $\mu$ is the most 
constrained SUSY CP phase, as suggested by  EDM bounds, 
a measurement of the normal tau polarisation will be a powerful 
tool to constrain  $\varphi_\mu$ independently from the low energy
measurements.  
Moreover, we have shown that the asymmetry can be sizable 
in inverted hierarchy scenarios, with  
heavy sfermions of the first and second generations, 
where the EDM constraints on the SUSY phases are less severe.

\medskip

To summarise, CP asymmetries in the decay of a chargino into a
polarised tau are one of the most sensitive probes to measure or
constrain $\varphi_\mu$ at the ILC. Since the feasibility of measuring
the tau polarisation can only be addressed in a detailed experimental
study, we want to motivate such thorough analyses, to explore the
potential of measuring SUSY CP phases at high energy colliders.

\section*{Acknowledgements}\label{sec:acknowledge}
We thank F.~von~der~Pahlen for enlightening discussions and helpful
comments.  This work has been partially funded by MICINN project
FPA.2006-05294.  AM was supported by the \textsl{Konrad Adenauer
  Stiftung}, BCGS and a fellowship of Bonn University.  HD and OK
thank the Aspen Centre for Physics for hospitality.  The authors are
grateful for financial support of the Ministerium f\"ur Innovation,
Wissenschaft, Forschung und Technologie (BMBF) Germany 
under grant numbers 05HT6PDA, and 05H09PDE.

\appendix

\section{Momenta and spin vectors}\label{sec:momenta}
We choose a coordinate frame for the centre-of-mass system such that
the momentum ${\bf{p}}_{{\tilde\chi}_j}$ of the chargino ${\tilde\chi}
_j$ points in the $z$-direction~\cite{MoortgatPick:1998sk}. The
scattering angle $\varangle({\bf{p}}_{e^-},{\bf{p}}_{{\tilde
\chi}_j})$ is denoted by $\theta$, and the azimuthal angle is set to
zero. Explicitly the momenta are then~\cite{MoortgatPick:1998sk}
\begin{eqnarray}  
p_{e^-}^\mu &=& E_b(1,-\sin\theta,0,\cos\theta),\\  
p_{{\tilde\chi}_i}^\mu&=&(E_{{\tilde\chi}_i},0,0,-q),\\   
p_{e^+}^\mu &=& E_b(1,\sin\theta,0,-\cos\theta),\\  
p_{{\tilde\chi}_j}^\mu &=&(E_{{\tilde\chi}_j},0,0,q),     
\end{eqnarray}  
with the beam energy $E_b=\sqrt{s}/2$, and~\cite{MoortgatPick:1998sk}  
\begin{eqnarray}  
E_{{\tilde\chi}_i}&=&\frac{s+m_{{\tilde\chi}_i}^2-m_{{\tilde\chi}_j}^2}  
                          {2\sqrt{s}},\\  
E_{{\tilde\chi}_j}&=&\frac{s+m_{{\tilde\chi}_j}^2-m_{{\tilde\chi}_i}^2}  
                          {2\sqrt{s}},\\  
q&=&\frac{\sqrt{\lambda (s,m_{{\tilde\chi}_i}^2,m_{{\tilde\chi}_j}^2)}}  
       {2\sqrt{s}},  
\end{eqnarray}  
with 
\begin{equation}  
\lambda(x,y,z)=x^2 + y^2+ z^2 -2(xy+xz+yz).  
\label{lambda}  
\end{equation}  
For the  chargino decay, 
${\tilde\chi}_i^\pm\to\tau^\pm{\tilde\nu}_\tau^{(\ast)}$, 
we parametrise the tau momentum in terms of the decay angle  
$\theta_D=\varangle({\bf{p}}_{\tau},{\bf{p}}_{{\tilde\chi}_i})$, and its  
azimuth $\phi_D$, 
\begin{eqnarray}  
  \left(p_{\tau}^\mu\right)^T&=&\left(\begin{array}{c}  
                                E_{\tau}\\  
                               -|{\bf{p}}_{\tau}|\sin\theta_D\cos\phi_D\\  
                                |{\bf{p}}_{\tau}|\sin\theta_D\sin\phi_D\\  
                               -|{\bf{p}}_{\tau}|\cos\theta_D  
                                \end{array}\right),\\[3mm]  
          |{\bf{p}}_{\tau}|    &=&  
                  \frac{m_{{\tilde\chi}_i}^2-m_{{\tilde\nu}_\tau}^2}  
                       {2(E_{{\tilde\chi}_i}-q\cos\theta_D)}.  
\end{eqnarray}  
The $\tau$ spin vectors are defined as  
\begin{eqnarray}  
s_{\tau}^{1,\mu}&=&\left(0,\frac{{\bf{s}}_{\tau}^2\times  
                     {\bf{s}}_{\tau}^3}{\left|{\bf{s}}_{\tau}^2  
                     \times{\bf{s}}_{\tau}^3\right|}\right),  
                     \label{tau1spin}\\[2mm]   
s_{\tau}^{2,\mu}&=&\left(0,\frac{{\bf{p}}_{\tau}\times{\bf{p}}_{e^-}}  
                     {\left|{\bf{p}}_{\tau}  
                    \times{\bf{p}}_{e^-}\right|}\right),\label{tauspin}\\[2mm]  
s_{\tau}^{3,\mu}&=&\frac{1}{m_{\tau}}  
                      \left(|{\bf{p}}_{\tau}|,  
                      \frac{E_{\tau}}{|{\bf{p}}_{\tau}|}  
                       {\bf{p}}_{\tau}\right).  
\label{tau3spin}  
\end{eqnarray}  
with 
\begin{equation}  
s^a_{\tau}\cdot s^b_{\tau} = -\delta^{ab},\quad  
s^a_{\tau}\cdot p_{\tau} = 0.  
\end{equation}  
The chargino and tau spin vectors fulfil the completeness 
relation~\cite{Haber:1994pe,MoortgatPick:1998sk}  
\begin{equation} 
\sum_c s_{\tilde\chi_k}^{c,\mu}\cdot s_{\tilde\chi_k}^{c,\nu} =-g^{\mu\nu} + 
\dfrac{p_{\tilde\chi_k}^\mu p_{\tilde\chi_k}^\nu}{m_{\tilde\chi_k}^2}. 
\label{completeness} 
\end{equation} 
  
\section{Phase space}\label{sec:phasespace}
For chargino production $e^+ e^-\to {\tilde\chi}_i^\pm {\tilde\chi}_j^\mp$, 
and  subsequent decay of one of the charginos,  
${\tilde\chi}_i^\pm\to\tau^\pm {\tilde\nu}_\tau^{(\ast)}$,  the  
Lorentz invariant phase-space element decomposes into two-body phase-space  
elements~\cite{Bycki} 
\begin{eqnarray}  
\dLips(s;p_{{\tilde\chi}_j},p_{\tau},p_{{\tilde\nu}_\tau})\!&=&  
\!\frac{1}{2\pi}  
\dLips(s;p_{{\tilde\chi}_i},p_{{\tilde\chi}_j})  
{\rm d}s_{{\tilde\chi}_i}\quad\nonumber\\  
\!&&\!\times\dLips(s_{{\tilde\chi}_i};p_{\tau},p_{{\tilde\nu}_\tau})\,.  
\end{eqnarray}  
The decay of the other chargino ${\tilde\chi}_j^\mp$ is not considered further. 
The constituent parts are 
\begin{eqnarray}  
\dLips(s;p_{{\tilde\chi}_i},p_{{\tilde\chi}_j})\!&=&\!  
              \frac{q}{8\pi\sqrt{s}}\sin\theta\,{\rm d}\theta,\\  
\dLips(s_{{\tilde\chi}_i};p_{\tau},p_{{\tilde\nu}_\tau})  
\!&=&\!        \frac{1}{2(2\pi)^2}  
               \frac{|{\bf p}_{\tau}|^2}  
                    {m_{{\tilde\chi}_i}^2 -m_{{\tilde\nu}_\tau}^2}\,
               {\rm d}\Omega_D,\qquad\enspace  
\end{eqnarray}  
with ${\rm d}\Omega_D = \sin\theta_D {\rm d}\theta_D {\rm d}\phi_D$, 
and $s_{{\tilde\chi}_i}=p_{{\tilde\chi}_i}^2$. \\ 
 
We use the narrow width approximation for the chargino propagator, 
 $\Delta({\tilde\chi}_i)$, Eq.~\eqref{propagator},  
\begin{equation} 
\int |\Delta({\tilde\chi}_i)|^2 \,{\rm d}s_{{\tilde\chi}_i}=  
\frac{\pi}{m_{{\tilde\chi}_i}\Gamma_{{\tilde\chi}_i}}\,. 
\label{nwapprox} 
\end{equation} 
This approximation should be justified for  
$(\Gamma_{{\tilde\chi}_i}/m_{{\tilde\chi}_i})^2\ll 1$, 
which holds in our case for chargino widths  
$\Gamma_{{\tilde\chi}_i}\lsim1~{\rm GeV}$ and masses  
$m_{{\tilde\chi}_i}\approx100~{\rm GeV}$. However, 
the naive $\mathcal{O}(\Gamma/m)$-expectation of the error can easily receive 
large  off-shell corrections of an order of magnitude and more, in particular 
at threshold, or due to interferences with other resonant or non-resonant  
processes. For a recent discussion of these issues, see, for example,  
Ref.~\cite{narrowwidth}.  
 
\section{Chargino diagonalisation matrices}\label{sec:parametrisation}
The matrices $U$ and $V$, which diagonalise the chargino  
matrix $M_{\tilde\chi}$, see Eq.~\eqref{eq:charmassmatrix}, 
can be parametrised by~\cite{Bartl:2002uy} 
\begin{eqnarray}  
U=  
\left(\begin{array}{cc}  
      e^{i\gamma_1} & 0\\  
      0             & e^{i\gamma_2}  
      \end{array}\right)  
\left(\begin{array}{cc}  
      \cos\theta_1              & e^{i\phi_1}\sin\theta_1\\  
      -e^{-i\phi_1}\sin\theta_1 & \cos\theta_1  
      \end{array}\right),  
\label{U}  
\end{eqnarray}  
\begin{eqnarray}  
V=  
\left(\begin{array}{cc}  
      \cos\theta_2             & e^{-i\phi_2}\sin\theta_2\\  
      -e^{i\phi_2}\sin\theta_2 & \cos\theta_2  
      \end{array}\right). 
\label{V}  
\end{eqnarray}  
The mixing angles, $-\pi/2\leq\theta_{1,2}\leq 0$, are  
\begin{subequations}\label{uvangles1}  
\begin{align}  
\frac{t(2\theta_1)}{2\sqrt{2}m_W} &=\frac{\sqrt{M_2^2 c^2(\beta)  
                                           +|\mu|^2 s^2(\beta)  
                                           +M_2|\mu|s(2\beta)c(\varphi_\mu)}}  
                                     {M_2^2 - |\mu|^2 -2m_W^2c(2\beta)},\\  
\frac{t(2\theta_2)}{2\sqrt{2}m_W} &=\frac{\sqrt{M_2^2 s^2(\beta)  
                                          +|\mu|^2 c^2(\beta)  
                                          +M_2|\mu| s(2\beta)c(\varphi_\mu)}}  
                                     {M_2^2 - |\mu|^2 +2m_W^2 c(2\beta)}, 
\end{align}  
\end{subequations} 
with the short hand notations $s(\alpha)=\sin(\alpha)$, 
$c(\alpha)=\cos(\alpha)$,  
and  $t(\alpha)=\tan(\alpha)$. 
For a CP-violating chargino system, $\varphi_\mu \neq 0$, 
the following CP phases enter 
\begin{subequations}\label{uvangles2} 
\begin{align} 
t(\phi_1) &=\phantom{-}s(\varphi_\mu)  
            \left[c(\varphi_\mu)+\frac{M_2}{|\mu|t(\beta)}\right]^{-1},\\  
t(\phi_2) &=-s(\varphi_\mu)  
            \left[c(\varphi_\mu)+\frac{M_2t(\beta)}{|\mu|}\right]^{-1},\\  
t(\gamma_1) &=- \dfrac{s(\varphi_\mu)}{  
            \left[c(\varphi_\mu)+\dfrac{M_2\left(m_{\tilde{\chi}_{1}}^2-  
                       |\mu|^2\right)}{|\mu|m_W^2 
            s(2\beta)}\right]},\label{gamma1}\\   
t(\gamma_2) &=\phantom{-}\dfrac{s(\varphi_\mu)}{  
            \left[c(\varphi_\mu)+\dfrac{M_2m_W^2 s(2\beta)}  
                       {|\mu|\left(m_{\tilde{\chi}_{2}}^2 -M_2^2\right)}   
                                        \right]} \,. 
\end{align}  
\end{subequations}  
The chargino masses are   
\begin{equation}  
m^2_{\tilde{\chi}_{1,2}}=\frac{1}{2} 
                          \biggl(M_2^2 +|\mu|^2+2m_W^2 \mp\kappa\biggr), 
\end{equation} 
with 

\begin{eqnarray} 
\kappa^2&=&\left(M_2^2-|\mu|^2\right)^2+4m_W^4c^2({2\beta})\nonumber\\  
        & &     +4m_W^2\left[M_2^2 +|\mu|^2  
                             +2M_2|\mu|s({2\beta})c(\varphi_\mu)\right].\qquad 
\end{eqnarray}

\section{Theoretical statistical significance}\label{sec:significance}
To measure a non-zero value of the CP asymmetry $\mathcal{A}_\tau^{\rm CP}$, 
Eq.~(\ref{asym}), over statistical fluctuations, we  define its 
theoretical statistical significance by~\cite{Trans} 
\begin{equation}  
S_\tau=\frac{|\mathcal{A}_\tau^{\rm CP}|}{\sigma_{\mathcal{A}}},  
\label{significanceA}  
\end{equation}   
such that $S_\tau$ is the expected number of standard  
deviations $\sigma_\mathcal{A}$ to which the asymmetry  
$\mathcal{A}_\tau^{\rm CP}$ can be determined to differ  
from zero. 
Since the variance is given by   
\begin{equation}  
\sigma_\mathcal{A}^2=\frac{1-|{\mathcal{A}_\tau^{\rm CP}}|^2}{2N},  
\end{equation}   
with the total number of events $N= \mathcal{L} \sigma$, we find 
\begin{equation}  
S_\tau=\frac{|\mathcal{A}_\tau^{\rm CP}|\sqrt{2\mathcal{L}\sigma}}  
            {\sqrt{1-|{\mathcal{A}_\tau^{\rm CP}}|^2}}\,.  
\label{significance} 
\end{equation}  
Note that our definition of the statistical significance $S_\tau$  
is purely based on the theoretical signal rate and its asymmetry.  
Detector efficiencies, event reconstruction efficiencies, and contributions  
from CP-even backgrounds are neglected, which would reduce the significance.  
The definition has thus to be regarded as an absolute upper bound only.  
In order to give realistic values of the statistical significances  
to observe a CP signal, a detailed experimental study is necessary.

  
\end{document}